\documentclass[pra,twocolumn,preprintnumbers,superscriptaddress]{revtex4}
\usepackage{amssymb}
\usepackage{amsmath}
\usepackage{amsfonts}
\usepackage{wasysym}
\usepackage{graphicx}
\usepackage{dcolumn}
\usepackage{bm}
\usepackage{color}

\begin{document}

\title{Quantum Phases of Self-Bound Droplets of Bose-Bose Mixtures}
\date{\today }

\author{Junqiao Pan}
\affiliation{CAS Key Laboratory of Theoretical Physics, Institute of Theoretical Physics, Chinese Academy of Sciences, Beijing 100190, China}
\affiliation{School of Physical Sciences, University of Chinese Academy of Sciences, Beijing 100049, China}

\author{Su Yi}
\email{syi@itp.ac.cn}
\affiliation{CAS Key Laboratory of Theoretical Physics, Institute of Theoretical Physics, Chinese Academy of Sciences, Beijing 100190, China}
\affiliation{School of Physical Sciences, University of Chinese Academy of Sciences, Beijing 100049, China}
\affiliation{CAS Center for Excellence in Topological Quantum Computation, University of Chinese Academy of Sciences, Beijing 100049, China}
\affiliation{Shenzhen Institute for Quantum Science and Engineering, Southern University of Science and Technology, Shenzhen 518055, China}

\author{Tao Shi}
\email{tshi@itp.ac.cn}
\affiliation{CAS Key Laboratory of Theoretical Physics, Institute of Theoretical Physics, Chinese Academy of Sciences, Beijing 100190, China}
\affiliation{CAS Center for Excellence in Topological Quantum Computation, University of Chinese Academy of Sciences, Beijing 100049, China}

\begin{abstract}
We systematically investigate the ground-state properties of self-bound droplets of quasi-two-dimensional binary Bose gases by using the Gaussian state theory. We find that quantum droplets consists two macroscopic squeezed phases and a macroscopic coherent phase. We map out the phase diagram and determine all phase boundaries via both numerical and nearly analytical methods. In particular, we find three easily accessible signatures for the quantum phases and the stablization mechanism of the self-bound droplets by precisely measuring their radial size. Our studies indicate that binary droplets represent an ideal platform for in-depth investigations of the quantum nature of the droplet state.
\end{abstract}

\maketitle

{\it Introduction}.---Quantum droplets in both dipolar~\cite{Rosensweig,singledropDy,singledropEr,beyondQF} and binary condensates~\cite{Cabrera,2dropsoli-drop,2drop3d,hetero2019,hetero2020} represent a new state of matter emerging in the mean-field unstable regime. They have attracted great interests in studying their fascinating properties, such as self-bound feature~\cite{singledropDy,singledropEr,Cabrera,2drop3d}, collective excitations~\cite{excisupersolidEr,excisupersolidDy}, soliton to droplet transition~\cite{2dropsoli-drop}, droplet-droplet collision~\cite{2dropcoll}, supersolid states~\cite{supersolid1,supersolid2,supersolid3}, Goldstone mode~\cite{Goldstone}, and so forth (see recent reviews~\cite{Boris_Review,Pfau_Review_long,Pfau_Review_short} and references therein). A widely adopted treatment for droplet states is the extend Gross-Pitaevskii equation (EGPE) which perturbatively incorporates quantum fluctuation into the Gross-Pitaevskii equation in terms of the Lee-Huang-Yang correction~\cite{LHY,dipolarTh1,dipolarTh2,dipolarTh3,twocomponentTh}. Although EGPE has provided satisfactory explanations to experimental observations, there are still discrepancies with experimental measurements~\cite{Cabrera,beyondQF,Pfau_Review_long}. 

Theoretical methods beyond EGPE have also been employed to study different aspects of the droplet states~\cite{QMC1, QMC2, QMC3, QMC4, QMC5, cbogoliubovss, Qcorrelations, bosonicp1, bosonicp2, Beliaev}. Among them, we studied the dipolar droplets using the Gaussian-state theory (GST) in which quantum fluctuation is included in a self-consistent manner~\cite{Wang2020}. Compared to other theoretical approaches beyond the EGPE theory, GST is computationally efficient and can be applied to realistic systems with large number of atoms. Interestingly, we found that, besides coherent state, dipolar droplets also contained two new macroscopic squeezed phases which were characterized by large second-order correlation and asymmetric atom-number distribution (AND), in striking contrast with those of the coherent-state-based droplets~\cite{Wang2020,Shi2019}. Whereas the asymmetric AND was experimentally observed~\cite{singledropDy,beyondQF}, verifying the predication on the second-order correlation function requires further experiments beyond simple density measurements.
	
In this Letter, we study the quantum phases of self-bound droplets in quasi-two-dimensional (quasi-2D) binary condensates. Although this system has same quantum phases as those in dipolar droplets, here we discover three readily accessible signatures for the determination of the quantum states and the stablization mechanisms. Specifically, due to the multiple quantum phases containing in the droplet states, the radial size ($\sigma$) versus atom number ($N$) curve is of W shape, as opposed to the V-shaped curve expected for single quantum phase. Therefore, the observation of double dips on the $\sigma$-$N$ curve may rule out the pure coherent explanation of the droplet state. Moreover, the {\em critical atom number} (CAN), i.e., the minimal number of atoms required for forming a self-bound state, is determined by the quantum states of the condensate and its precise measurement of the CAN allows us to distinguish squeezed state from coherent one. Finally,  the {\em dip atom number} (DAN), i.e., the atom number at the dip of the $\sigma$-$N$ curve, depends on both quantum phase and stabilization mechanism, which makes it a quantitative criterion for stability mechanism.

{\it Formulation}.---We consider a ultracold gas of $N$ ${}^{39}\text{K}$ atoms with $N_{\uparrow}$ atoms being in the hyperfine state $\left|\uparrow \right\rangle \equiv \left| F, m_{F} \right\rangle = \left| 1, -1 \right\rangle$ and $N_{\downarrow}$ in $\left| \downarrow \right\rangle \equiv \left| 1, 0 \right\rangle$. The total Hamiltonian of the system, 
$$H = H_{0} + H_{2B} + H_{3B},$$ 
consists of the single-, two-, and three-particle terms. In second-quantized form, the single-particle part reads 
\begin{align}
H_{0} = \sum_{\alpha} \int d \mathbf{r} \hat{\psi}^{\dagger}_{\alpha} ( \mathbf{r} ) h_{\alpha} \hat{\psi}_{\alpha}( \mathbf{r} ),
\end{align}
where $\hat{\psi}_{\alpha}\left( \mathbf{r} \right)$ ($\alpha=\uparrow,\downarrow$) are the field operators and $h_{\alpha} = - \hbar^2\nabla^{2}/ ( 2 M ) - \mu_{\alpha}$ with $M$ being the mass of the atoms and $\mu_{\alpha}$ the chemical potential introduced to fixed the atom number in the $\alpha$th component. The two-body (2B) interaction Hamiltonian takes the form
\begin{equation}
	H_{2B} = \sum_{\alpha\beta} \frac{g_{\alpha\beta}}{2} \int d \mathbf{r}  \hat{\psi}^{\dagger}_{\alpha}( \mathbf{r} ) \hat{\psi}^{\dagger}_{\beta}( \mathbf{r} )  \hat{\psi}_{\beta}( \mathbf{r} ) \hat{\psi}_{\alpha}( \mathbf{r} ),
\end{equation}
where $g_{\alpha\beta} = 4 \pi\hbar^2 a_{\alpha\beta} / M$ characterize the 2B interaction strengths with $a_{\alpha\beta}$ being scattering length between components $\alpha$ and $\beta$. The scattering lengths are tunable through Feshbach resonance and for scenario of interest to quantum droplets, the intra- and inter-species scattering lengths satisfy $a_{\uparrow\uparrow},a_{\downarrow\downarrow}>0$ and $a_{\uparrow\downarrow}<0$, respectively. Finally, the three-body (3B) interaction Hamiltonian can be expressed as
\begin{align}
	H_{3B} = \frac{g_{3}}{3!} \sum_{\alpha\beta\gamma} \int d \mathbf{r} \hat{\psi}^{\dagger}_{\alpha}( \mathbf{r} ) \hat{\psi}^{\dagger}_{\beta}( \mathbf{r} ) \hat{\psi}^{\dagger}_{\gamma}( \mathbf{r} )\hat{\psi}_{\gamma}( \mathbf{r} ) \hat{\psi}_{\beta}( \mathbf{r} ) \hat{\psi}_{\alpha}( \mathbf{r} ),
\end{align}
where $g_{3}$ is the 3B coupling constant which, for simplicity, is assumed to be independent of the spin components. Moreover, since we are only interested in the ground states of the system, we assume that $g_3$ is real and positive. The value of $g_3$ shall be determined by fitting the experimental data.

We study self-bound droplet states using the GST. Specifically, we assume that the many-body wave function of a binary gas adopts the variational ansatz~\cite{Shi2018,Shi2019,Tommaso}
\begin{equation}
	| \Psi_{\rm GS} \rangle = e^{\hat{\Psi}^{\dagger}( \mathbf{r} ) \Sigma^{z}( \mathbf{r}, \mathbf{r}^{\prime} ) \Phi( \mathbf{r}^{\prime} )} e^{i \hat{\Psi}^{\dagger}( \mathbf{r} ) \xi( \mathbf{r}, \mathbf{r}^{\prime} ) \hat{\Psi}( \mathbf{r}^{\prime} )/2} | 0 \rangle, \label{GS}
\end{equation}
where $\hat{\Psi}( \mathbf{r} ) \equiv \begin{pmatrix}
\hat{\psi}( \mathbf{r} )\\  \hat{\psi}^{\dagger}( \mathbf{r} )
\end{pmatrix}$ with $\hat{\psi}( \mathbf{r} ) = \begin{pmatrix}
\hat{\psi}_{\uparrow}( \mathbf{r} )\\ \hat{\psi}_{\downarrow}( \mathbf{r} )
\end{pmatrix}$ is the field operator expressed in the Nambu basis, 
\begin{align}
\Sigma^z( \mathbf{r}, \mathbf{r}^{\prime} )=\begin{pmatrix}
I_2\otimes\delta( \mathbf{r}- \mathbf{r}^{\prime} )&0\\0&-I_2\otimes\delta( \mathbf{r}- \mathbf{r}^{\prime} )
\end{pmatrix}
\end{align}
with $I_2$ being a $2\times2$ identity matrix in the spin space, $\Phi( \mathbf{r} ) = \langle \hat{\Psi}( \mathbf{r} ) \rangle = \begin{pmatrix}
\phi^{(c)}\\ \phi^{(c)*}\end{pmatrix}$ with $\phi^{(c)} =\begin{pmatrix}
 \phi^{(c)}_{\uparrow}\\ \phi^{(c)}_{\downarrow}
\end{pmatrix}$ is the variational parameter describing the coherent part of the condensates, and $\xi( \mathbf{r}, \mathbf{r}^{\prime} )$ is the variational parameter that characterizes the squeezed part of the condensate. The ground state wave function can be found by minimizing energy $E_{0} = \left\langle \Psi_{\rm GS} \right| H \left| \Psi_{\rm GS} \right\rangle$ through imaginary-time evolution~\cite{Shi2018,Shi2019,SM}.

Physically, it is more convenient to factorize the Gaussian state wave function into a multimode squeezed coherent state~\cite{Wang2020,SM}, i.e.,
\begin{equation}
	\left| \Psi_{\rm GS} \right\rangle = e^{\sqrt{N^{(c)}}\left( \hat{c}^{\dagger} - \hat{c}^{} \right)} \prod^{\infty}_{i = 1} e^{\frac{1}{2} \xi_{i} \left( \hat{s}^{\dagger 2}_{i} - \hat{s}^{2}_{i} \right)} \left| 0 \right\rangle, \label{ground state wave function}
\end{equation}
where $N^{(c)}=\int d \mathbf{r}\left[\phi^{(c)}(\mathbf{r})\right]^\dag \phi^{(c)}(\mathbf{r})$ is the atom number in the coherent mode and $\hat{c}=\int d \mathbf{r}  \left[\bar{\phi}^{(c)}(\mathbf{r})\right]^\dag\hat{\psi}(\mathbf{r})$ with $\bar{\phi}^{(c)}(\mathbf{r})=\phi^{(c)}(\mathbf{r}) / \sqrt{N^{(c)}}$
being the normalized mode function for the coherent component. In the squeezing operators, $\hat s_i=\int d \mathbf{r} \left[\bar{\phi}_{i}^{(s)}(\mathbf{r})\right]^\dag\hat{\psi}(\mathbf{r})$ is the annihilation operator for the $i$th squeezed mode $\bar{\phi}^{(s)}_{i}\equiv\begin{pmatrix}
{\phi}^{(s)}_{i,\uparrow}\\{\phi}^{(s)}_{i,\downarrow}
\end{pmatrix}$ that satisfy $\int d \mathbf{r} \left[\bar{\phi}_{i}^{(s)}({\mathbf r})\right]^\dag \bar{\phi}^{(s)}_{j}({\mathbf r})=\delta_{i j}$. Furthermore, $\sinh \xi_{i}=\sqrt{N^{(s)}_{i}}$ with $N_i^{(s)}$ being the occupation number in the $i$th squeezed mode. The total number of atoms in the squeezed modes is $N^{(s)}=\sum_{i} N^{(s)}_{i}$ and the total squeezed density is $n^{(s)}({\mathbf r})=\sum_iN_i^{(s)}|\bar{\phi}_{i}^{(s)}({\mathbf r})|^2$. For convenience, it is always assumed that $N^{(s)}_{i}$ are sorted in descending order with respect to the index $i$. Mode $j$ is notably populated if $N^{(s)}_{j} / N^{(s)} \geq 0.1\%$~\cite{Wang2020}. 

In the main text, we focus on the self-bound droplets in quasi-2D geometry achieved by imposing a harmonic confinement, $V_z(z)=M\omega_z^2z^2/2$, along the $z$ axis~\cite{Cabrera}. When $\omega_z$ is sufficiently large, the motion of atoms along the $z$ direction is frozen to the ground state of $V_z(z)$. In additioin, we always fix the atom number ratio between two components at $N_\uparrow/N_\downarrow=\sqrt{a_{\downarrow\downarrow}/a_{\uparrow\uparrow}}$ by following the experiments~\cite{Cabrera,2drop3d}. Numerical results can be conveniently checked using the virial relation, $E_{k} + E_{2B} + 2 E_{3B} = 0$, where $E_{k}$, $E_{2B}$,  and $E_{3B}$ are kinetic, 2B, and 3B energies, respectively~\cite{SM}. We point out that results for 3D droplets are presented in the Supplemental Material~\cite{SM}.

\begin{figure}[ptb]
\centering
\includegraphics[width=0.95\columnwidth]{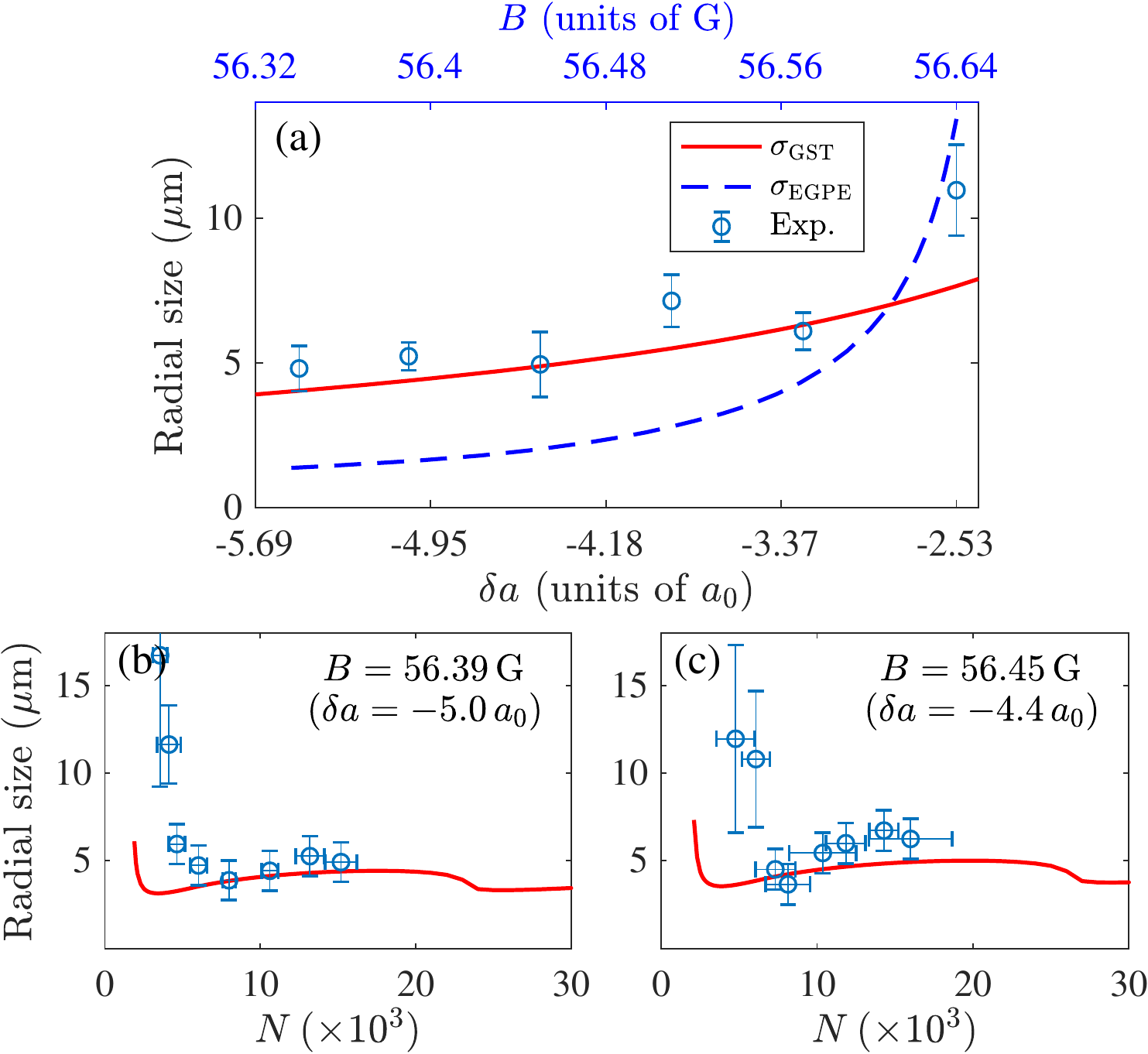}
\caption{(color online). (a) Radial size versus reduced scattering length for $N = 1.5 \times 10^{4}$. Solid line is computed using the GST with $g_3=6.65 \times 10^{-39}\hbar {\rm m}^{6}/{\rm s}$. Empty circles with error bars and dashed line (both are extracted from Ref.~\cite{Cabrera}) represent the experimental data and the EGPE result, respectively. (b) and (c) show radial size as functions of atom number for various magnetic fields. Empty circles (extracted from Ref.~\cite{Cabrera}) are experimental data. More comparisons between the numerically computed radial size and the experimental data can be found in the Supplemental Material~\cite{SM}.}
\label{check_g3}
\end{figure}
 
{\it Three-body coupling strength}.---Let us first fix the value of $g_3$ for K atom. Previously, $g_3$ of Dy atom was determined by fitting the AND of the droplet states~\cite{Wang2020}. Here we instead determine $g_3$ by fitting the radial size $\sigma$ of the quasi-2D droplets. Following the experiment~\cite{Cabrera}, we extract $\sigma$ by fitting the total density with a two-dimensional Gaussian. In Fig.~\ref{check_g3}(a), we plot the radial size $\sigma$ as a function of the effective scattering length $\delta a\equiv a_{\uparrow\downarrow} +\sqrt{a_{\uparrow\uparrow}a_{\downarrow\downarrow}}$ (or, equivalently, the magnetic field $B$) computed with $g_{3} = 6.65 \times 10^{-39}\hbar {\rm m}^{6}/{\rm s}$. Good agreement between the numerical and the experimental results is achieved. We point out that varying the value of $g_3$ does not change the shape of the red line; it only shifts the red line vertically. Therefore, to achieve better agreement for the fitting, one may have to make $g_3$ spin dependent. 

To further demonstrate the validity of the fitted $g_3$, we show, in Fig.~\ref{check_g3}(b) and (c), the $N$ dependence of $\sigma_{\rm GST}$ under different magnetic fields. As can be seen, although discrepancies exist at small $N$, good agreements are still achieved for large $N$. Particularly, the main feature of the experimental results is captured by the numerical simulations (see below for details). Hereafter, we shall always use the fitted $g_3$ for all numerical results presented below.

\begin{figure}[ptb]
\centering
\includegraphics[width=1\columnwidth]{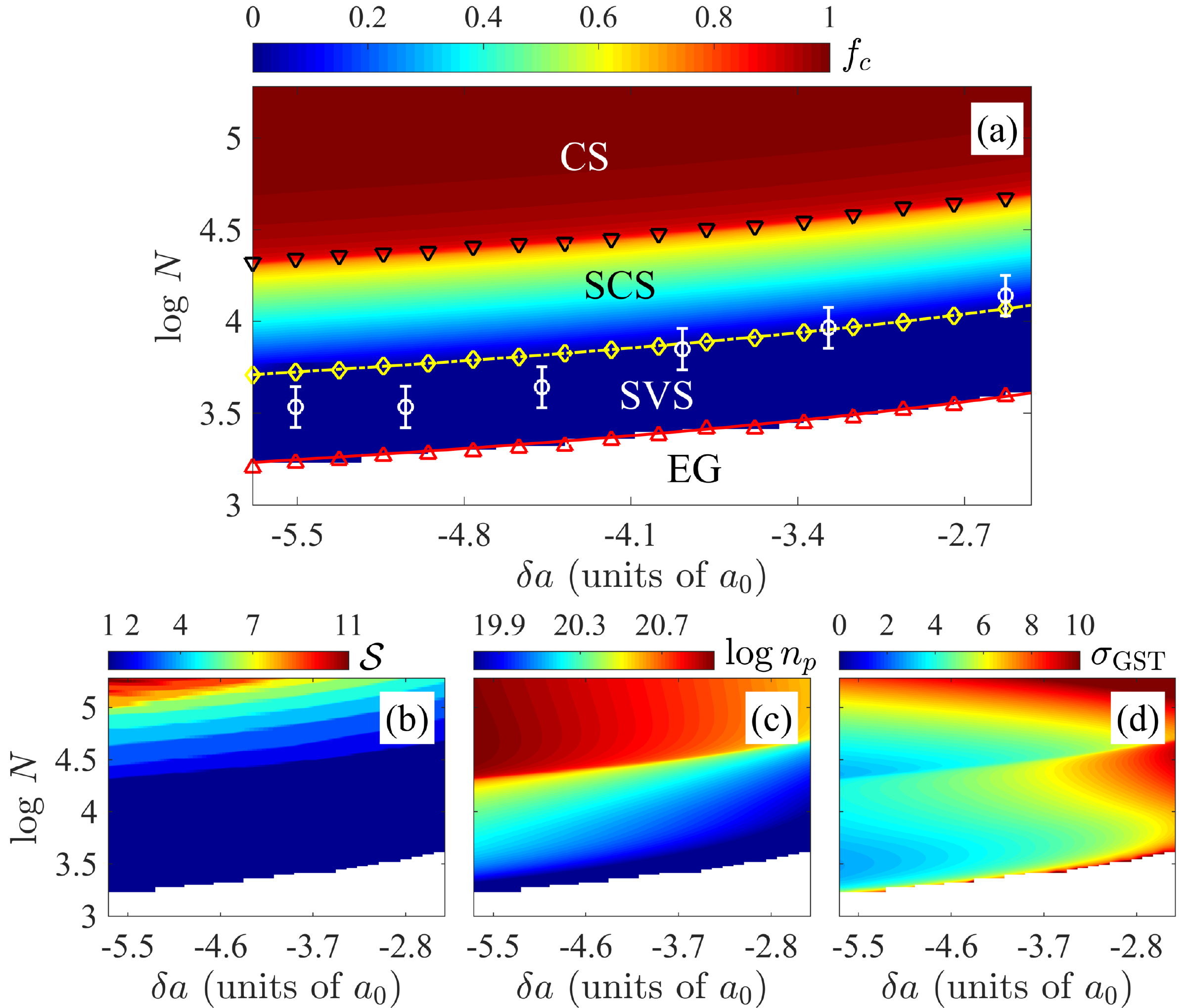}
\caption{(color online). Phase diagram. (a) Distribution of the coherent fraction $f_c$ on the $(\delta a,N)$ parameter plane. Markers $\triangle$, $\diamondsuit$, and $\bigtriangledown$ denote numerical results for CAN, SVS-to-SCS transition boundary, and SCS-to-CS transition boundary, respectively. The solid and dash-dotted lines show the corresponding analytic results. Empty circles ($\ocircle$) with error bar are experimental data for CAN~\cite{Cabrera}. (b)-(d) demonstrate the number of the notably-populated squeezed modes $\mathcal{S}$, the distribution of the peak density $n_p$ (units of ${\rm m}^{-3}$), and the radial size $\sigma_{\rm GST}$ (units of $\mu{\rm m}$) on the $(\delta a,N)$ parameter plane, respectively.}
\label{phase2d}
\end{figure}

{\it Phase diagram}.---Given that $N_{\uparrow}/N_{\downarrow}=\sqrt{a_{\downarrow\downarrow}/a_{\uparrow\uparrow}}$, it is found numerically that the coherent and squeezed modes satisfy $\phi^{(c)}_\uparrow({\mathbf r})/\phi^{(c)}_\downarrow({\mathbf r})=\sqrt{N_\uparrow/N_\downarrow}$ and  $\phi^{(s)}_{i,\uparrow}({\mathbf r})/\phi^{(s)}_{i,\downarrow}({\mathbf r})=\sqrt{N_\uparrow/N_\downarrow}$ for any $i$, respectively. As a result, $N^{(c)}_{\alpha}/N_{\alpha}$ is independent of the spin component $\alpha$, which allows us to define the coherent fraction as $f_{c} \equiv N^{(c)} / N$ to characterize the property of the condensate, instead of considering the coherent fraction of individual spin component. 

Figure~\ref{phase2d}(a) shows the distribution of $f_c$ on the $(\delta a,N)$ parameter plane. Similar to dipolar droplets~\cite{Wang2020}, there are four distinct regions: expanding gas (EG), squeezed-vacuum state (SVS), squeezed-coherent state (SCS), and coherent state (CS). The EG region lies below the CAN (solid line or $\triangle$), in which the attractive two-body interaction is insufficient for the gas to form a self-bound state. Both SVS and SCS phases contain a single squeezed mode except SCS phase also consists of a coherent component. The SVS-to-SCS transition (dash-dotted line or $\diamondsuit$) breaks the Z${}_2$ symmetry of the condensate and is a third-order phase transition. The CS phase is dominated by the coherent component with $f_c>0.75$ and the SCS-to-CS transition is of first order. Interestingly, as shown by the solid and the dash-dotted lines, the boundaries for the self-bound state and the SVS-to-SCS transition can be found nearly analytically with high accuracy~\cite{SM}. We may also determine the boundary of SCS-to-CS transition by analyzing the stability of Bogliubov excitation in the CS phase~\cite{SM}.

The phase diagram is most conveniently reproduced by plotting the number of the notably-populated squeezed modes $\mathcal{S}$. As shown in Fig.~\ref{phase2d}(b), both SVS and SCS phases are featured by single squeezed mode; while CS phase may contains a large number of squeezed modes even if the squeezed component becomes negligibly small. Physically, squeezing in SVS and SCS phases are macroscopic quantum state originating from two-body attraction~\cite{Shi2019,Wang2020}; while squeezing in CS phase represents quantum depletion induced by 3B repulsion~\cite{Wang2020}. Similar to dipolar droplets, the presence of squeezing also significantly modifies the AND of these quantum states~\cite{SM}. The properties of these quantum phases can be further explored by examining the peak density $n_p$ and the radial size $\sigma_{\rm GST}$ as plotted in Fig.~\ref{phase2d}(c) and (d), respectively. Among them, $\sigma_{\rm GST}$ exhibits visible difference as compared to the distributions of other quantities. 

\begin{figure}[ptb]
\centering
\includegraphics[width=0.75\columnwidth]{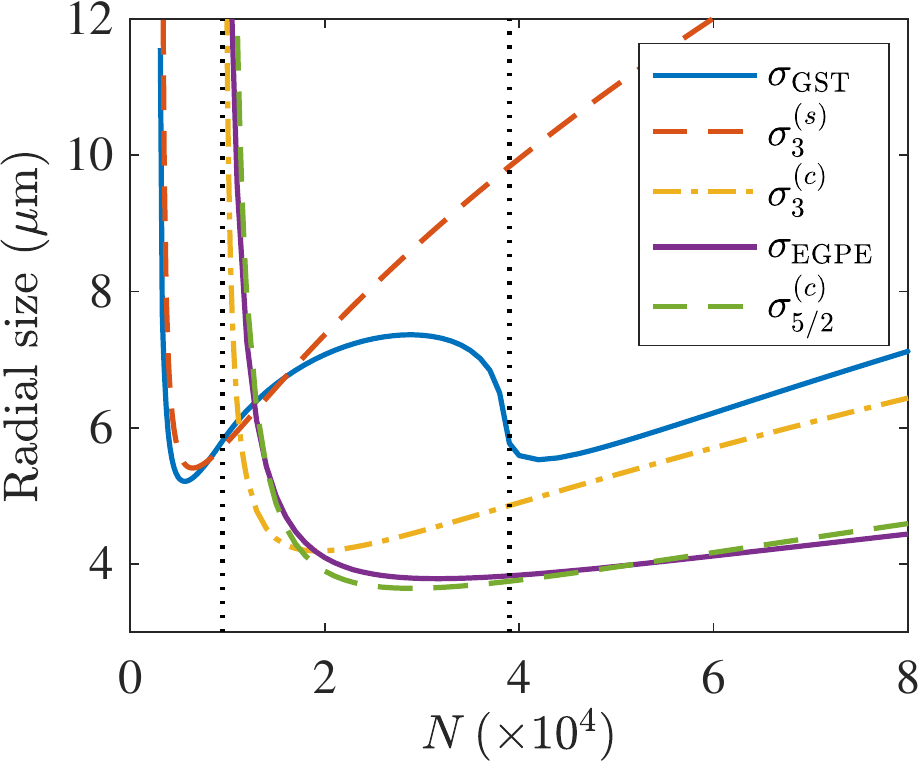}
\caption{(color online). Radial size versus mean atom number for $\delta a=-3.062 a_{0}$. Vertical dotted lines mark the locations of two phase transitions.}
	\label{size}
\end{figure}

{\em Radial size}.---To explore the properties of $\sigma_{\rm GST}$, we plot the $N$ dependence of $\sigma_{\rm GST}$ for $\delta a=-3.062 a_{0}$ in Fig.~\ref{size}. As a comparison, the radial size numerically computed via EGPE, $\sigma_{\rm EGPE}$, is also plotted. Surprisingly, $\sigma_{\rm GST}$ is W-shaped function of $N$, in striking contrast to the V-shaped $\sigma_{\rm EGPE}$. To understand its origin, we note that the total energy per atom can be expressed as~\cite{SM}
\begin{align}
\varepsilon_0=\frac{E_0(\sigma)}{N}\propto \frac{1}{\sigma^2}-\frac{\tilde g_2N}{\sigma^2}+\frac{\tilde g_\nu N^{\nu-1}}{\sigma^{2(\nu-1)}}\quad\mbox{(for $\nu>2$)},\label{engsig}
\end{align}
where $E_0=E_{k}+E_{2B}+E_{3B}$ are the total energy. On the right hand side, the first term is the kinetic energy, the second term is contributed by the 2B interaction, and the last term is the energy associated with the stabilization force, i.e., $\nu=3$ for 3B repulsion and $5/2$ for quantum fluctuation. Furthermore, $\tilde g_2\, (>0)$ and $\tilde g_\nu\, (>0)$ are reduced interaction parameters (RIPs). From Eq.~\eqref{engsig}, the equilibrium size can be immediately found to be
\begin{align}
\sigma=\sqrt{N}\left[\frac{(\nu-1)\tilde g_\nu N}{\tilde g_2N-1}\right]^{1/[2(\nu-2)]}\quad\mbox{(for $N>1/\tilde g_2$)},\label{sigman}
\end{align}
which is a V-shaped function of $N$. More specifically, $\sigma$ diverges as $N$ approaches the CAN 
\begin{align}
N_{\rm cri}\equiv \frac{1}{\tilde g_2},\label{ncri}
\end{align}
grows asymptotically as $\sigma\approx\sqrt{N}\left[(\nu-1)\tilde g_\nu/\tilde g_2\right]^{1/[2(\nu-2)]}$ for large $N$, and reaches its minimal value at the DAN
\begin{align}
N_{\rm dip}\equiv\frac{\nu-1}{(\nu-2)\tilde g_2}.\label{ndip}
\end{align}
In contrast, we note that the peak density, $n_p\propto N/\sigma^2=\left\{(\tilde g_2N-1)/[(\nu-1)\tilde g_\nu N]\right\}^{1/(\nu-2)}$, is always a monotonically increasing function of $N$ and converges to a constant proportional to $\left\{\tilde g_2/[(\nu-1)\tilde g_\nu ]\right\}^{1/(\nu-2)}$ for large $N$. It is worthwhile to point out that $\sigma(N)$ depends not only on the stabilization mechanism through $\nu$ but also on the many-body wave function via the RIPs $(\tilde g_2,\tilde g_\nu)$~\cite{SM}. 

To carry out quantitative comparisons, we assume that all density profiles are Gaussian functions. Then for pure squeezed and coherent states with 3B interaction, we obtain the RIPs $(\tilde g_2^{(s)},\tilde g_3^{(s)})$ and $(\tilde g_2^{(c)},\tilde g_3^{(c)})$, respectively, satisfying $\tilde g_2^{(s)}=3\tilde g_2^{(c)}$ and $\tilde g_3^{(s)}=15\tilde g_3^{(c)}$~\cite{SM,Wang2020}. In addition, for the coherent-state-based EGPE theory, the RIPs are $(\tilde g_2^{(c)},\tilde g_{5/2}^{(c)})$~\cite{SM}. Correspondingly, we plot $\sigma_\nu^{(c,s)}(N)$ using Eq.~\eqref{sigman} in Fig.~\ref{size}, where the subscript denotes the stabilization mechanism and the superscript denotes the quantum states. As can be seen, $\sigma_3^{(s)}$ agrees very well with $\sigma_{\rm GST}$ around the left dip where the condensate is a SVS. This agreement also indicates that the density profile of a SVS is well approximated by a Gaussian function. Around the right dip, the quantum state of the condensate is more complicated than a pure coherent state. Therefore, $\sigma_3^{(c)}$ only exhibits rough agreement with $\sigma_{\rm GST}$ for large $N$ where the quantum state is dominated by coherent component. Still, the remaining discrepancy can be explained as the density profile of the coherent state is significantly deviated from a Gaussian function~\cite{SM}. As to $\sigma_{5/2}^{({c})}$, it agree well with $\sigma_{\rm EGPE}$, which confirms the variational calculation. One should note that, although $\sigma_{5/2}^{(c)}$ and $\sigma_3^{(c)}$ have the same CANs, they define different DANs.

It is now clear that the W-shaped $\sigma_{\rm GST}$ stems from the multiple quantum phases in the self-bound droplets. Remarkably, this feature can be vaguely seen in the experimental data~\cite{Cabrera}. In fact, as shown in Fig.~\ref{check_g3}(b) and (c), after the completion of the left dip, the experimental data of the radial size starts to decrease again as one further increases $N$. Since radial size will eventually grow as $\sqrt{N}$ in the liquid phase, the drop after the first dip then signals the existence of the second dip and, consequently, the W shape. As the coherent-state-based EGPE theory only predicts a V-shaped radial size curve, a clear observation of the double dips may qualitatively exclude the EGPE explanation of the droplet states.

More interestingly, variational analyses can even lead to quantitative criteria for quantum phases and stabilization mechanisms. To see this, let us examine $N_{\rm cri}$, the CAN extracted analytically from Eq.~\eqref{sigman}. At first sight, it is surprised that $N_{\rm cri}$ is independent of $\tilde g_\nu$. This can be understood as follows. Because $\varepsilon_k$ and $\varepsilon_{2B}$ have the same power dependence on $\sigma$, the 2B attraction can be canceled out exactly by the kinetic energy when $N=N_{\rm cri}$. Then, in the absence of stabilization force, a self-bound state can be of arbitrary radial size. Once the stabilization force is turned on, in order to have a self-bound state of the same atom number, the radial size of the droplet has to be infinite to make the repulsive interaction energy vanish. This universality of $N_{\rm cri}$ is highly nontrivial as it allows our theory to be examined by experimental data regardless of the accuracy of the fitted $g_3$. We point out that the above analysis is not applicable in 3D geometry where a self-bound droplet always has a finite size~\cite{SM}.

Still, the CAN depends on the quantum state of the condensate through the RIP $\tilde g_2$. Then corresponding to the pure squeezed and coherent states, we have distinct CANs $N_{\rm cri}^{(s)}$ and $N_{\rm cri}^{(c)}$ which satisfy $N_{\rm cri}^{(c)}=3N_{\rm cri}^{(s)}$. The precise measurements of CAN can therefore be used to distinguish the quantum state of the condensates. In Fig.~\ref{phase2d}(a), we show the experimentally measured CAN versus the reduced scattering length. From the rightmost three data points, it seems that the quantum state at the self-bound boundary is coherent state; while the quantum state for the other three data point is unclear. However, after carefully examining these data, we find that high precision measurements are still needed to accurately determine the quantum phase~\cite{SM}. In fact, to precisely determine CAN, the experimental data should be fitted using proper curves corresponding to appropriate stabilization mechanisms. In particular, the experimental data should be of V shape close to the self-bound boundary.

Finally, we note that even the stabilization mechanism can be identified using DAN $N_{\rm dip}$, a quantity depending on $\nu$. From Eq.~\eqref{ndip}, it is clear that, corresponding to three $\sigma_\nu^{(c,s)}(N)$, there are three distinct DANs that satisfy  $N_{{\rm dip},3}^{(c)}=3N_{{\rm dip},3}^{(s)}$ and $N_{{\rm dip},5/2}^{(c)}=(9/2)N_{{\rm dip},3}^{(s)}$, where $N_{{\rm dip},3}^{(s)}=2/\tilde g_2^{(s)}$. Therefore, precise measurement of DAN allows us to distinguish not only the quantum phase but also the stabilization mechanism. Moreover, $N_{\rm dip}$ is also independent of $\tilde g_\nu$, which makes the determination of the stabilization mechanism independent of the fitting of $g_3$.

{\em Conclusion}.---We have provided a detailed phase diagram for self-bound droplets of quasi-2D binary condensates beyond the coherent-state based mean field theory. Among all quantum phases, SVS and SCS are two macroscopic quantum squeezed states that demand experimental verification. More importantly, other than measuring AND or second-order correlation function, we have found three easily accessible experimental signatures: i) the observation of the double dips on the $\sigma(N)$ curve rule out the explanations of the liquid droplets based on single quantum state; ii) the precise measurements of $N_{\rm cri}$ or $N_{\rm dip}$ further allow us to distinguish the squeezed and the coherent states; iii) we may even determine the stabilization mechanism from $N_{\rm dip}$. More remarkably, both $N_{\rm cri}$ and $N_{\rm dip}$ are independent of the strength of the stabilization force, which makes the comparison between the theoretical and the experimental results immune to any unknown parameters.

{\em Acknowledgement.}---We thank enlightening discussions with H. Hu, X.-J. Liu, C. R. Cabrera, C. Navarrete-Benlloch, T. Pfau, and H. P. Buechler. This work was supported by National Key Research and Development Program of China (Grant No. 2017YFA0304501), by the NSFC (Grants No. 11974363, and No. 12047503), by the Strategic Priority Research Program of CAS (Grant No. XDB28000000), and by the Open Project of Shenzhen Institute of Quantum Science and Engineering (Grant No. SIQSE202006).

\bibliography{ref_bindrop.bib}

%
%

\onecolumngrid
\clearpage

\begin{center}
\textbf{\large Supplemental Material}
\end{center}

\setcounter{equation}{0} 
\setcounter{figure}{0} 
\makeatletter

\renewcommand{\thefigure}{SM\arabic{figure}} \renewcommand{\thesection}{SM-\Roman{section}} \renewcommand{\theequation}{SM\arabic{equation}}

In the Supplemental Material, we present the derivation for the imaginary-time evolution in Gaussian-state theory (Sec.~\ref{secimaginary}), the atom number distribution of a Gaussian state (Sec.~\ref{secand}), the virial relation (Sec.~\ref{secvirial}), the density profile (Sec.~\ref{secdens}), the boundary of the SVS-to-SCS transition (Sec.~\ref{sec3rd}), the boundary of the SCS-to-CS transition (Sec.~\ref{sec1st}), the variational analysis for radial size (Sec.~\ref{secvariational}), and the main properties of the 3D droplets (Sec.~\ref{sec3ddrop}).

\section{Imaginary-time Evolution}\label{secimaginary}
Here we show the details of how to use the imaginary-time evolution to get the Gaussian mean-field ground state. To this end, we assume that the many-body wave function of a binary gas adopts the variational ansatz~\cite{Shi2018_SM,Shi2019_SM,Tommaso_SM,Wang2020_SM}
\begin{equation}
	| \Psi_{\rm GS} \rangle = e^{\hat{\Psi}^{\dagger}( \mathbf{r} ) \Sigma^{z}( \mathbf{r}, \mathbf{r}^{\prime} ) \Phi( \mathbf{r}^{\prime} )} e^{i \hat{\Psi}^{\dagger}( \mathbf{r} ) \xi( \mathbf{r}, \mathbf{r}^{\prime} ) \hat{\Psi}( \mathbf{r}^{\prime} )/2} | 0 \rangle, \label{dfGS}
\end{equation}
where $\hat{\Psi}( \mathbf{r} ) \equiv \begin{pmatrix}
	\hat{\psi}( \mathbf{r} )\\  \hat{\psi}^{\dagger}( \mathbf{r} )
\end{pmatrix}$ with $\hat{\psi}( \mathbf{r} ) = \begin{pmatrix}
	\hat{\psi}_{\uparrow}( \mathbf{r} )\\ \hat{\psi}_{\downarrow}( \mathbf{r} )
\end{pmatrix}$ is the field operator expressed in the Nambu basis, 
\begin{align}
	\Sigma^z( \mathbf{r}, \mathbf{r}^{\prime} )=\begin{pmatrix}
		I_2\otimes\delta( \mathbf{r}- \mathbf{r}^{\prime} )&0\\0&-I_2\otimes\delta( \mathbf{r}- \mathbf{r}^{\prime} )
	\end{pmatrix}
\end{align}
with $I_2$ being a $2\times2$ identity matrix in the spin space, $\Phi( \mathbf{r} ) = \langle \hat{\Psi}( \mathbf{r} ) \rangle = \begin{pmatrix}
	\phi^{(c)}\\ \phi^{(c)*}\end{pmatrix}$ with $\phi^{(c)} =\begin{pmatrix}
	\phi^{(c)}_{\uparrow}\\ \phi^{(c)}_{\downarrow}
\end{pmatrix}$ is the variational parameter describing the coherent part of the condensates, and $\xi( \mathbf{r}, \mathbf{r}^{\prime} )$ is the variational parameter that characterizes the squeezed part of the condensate. We point out that the multiplication in the exponents in Eq.~\eqref{dfGS} represent short-hand notations which can be computed by first performing the matrix multiplications in the Nambu space and then integrating over the repeated spatial coordinates. As an example, we have
\begin{align}
	\hat{\Psi}^{\dagger}( \mathbf{r} ) \Sigma^{z}(\mathbf{r}, \mathbf{r}^{\prime}) \Phi(\mathbf{r}^{\prime}) = \sum_{\alpha} \int d\mathbf{r} d\mathbf{r}^{\prime} \left[\hat{\psi}_{\alpha}^{\dagger}(\mathbf{r}) \delta(\mathbf{r} - \mathbf{r}^{\prime}) \phi^{(c)}_{\alpha}(\mathbf{r}^{\prime}) - \hat{\psi}_{\alpha}(\mathbf{r}) \delta(\mathbf{r} - \mathbf{r}^{\prime}) \phi^{(c) \ast}_{\alpha}(\mathbf{r}^{\prime}) \right].
\end{align}

It should be noted that there exists a gauge redundancy in $\xi\left(\mathbf{r}, \mathbf{r}^{\prime}\right)$~\cite{Shi2018_SM,Shi2019_SM}. Namely, different $\xi$ 's may lead to the same Gaussian state. To remove this redundancy, we introduce the covariance matrix
\begin{align}
	\Gamma( \mathbf{r}, \mathbf{r}^{\prime} )\equiv\begin{pmatrix}
		\Gamma_{11}( \mathbf{r}, \mathbf{r}^{\prime} )&\Gamma_{12}( \mathbf{r}, \mathbf{r}^{\prime} )\\
		\Gamma_{21}( \mathbf{r}, \mathbf{r}^{\prime} )&\Gamma_{22}( \mathbf{r}, \mathbf{r}^{\prime} )
	\end{pmatrix} = \langle \{ \delta \hat{\Psi}( \mathbf{r} ), \delta \hat{\Psi}^{\dagger}( \mathbf{r}^{\prime} ) \} \rangle,\nonumber
\end{align}
where $\delta \hat{\Psi} = \hat{\Psi} - \Phi$ is the fluctuation and $\{ ,  \}$ here denotes the anticommutator of the vectorial operator where in detail
\begin{align}
	\Gamma_{11}( \mathbf{r}, \mathbf{r}^{\prime} ) = \begin{pmatrix}
		\langle\{\delta\hat\psi^\dagger_{\uparrow}({\mathbf r}'), \delta\hat\psi_{\uparrow}({\mathbf r})\}\rangle & \langle\{\delta\hat\psi^\dagger_{\uparrow}({\mathbf r}'), \delta\hat\psi_{\downarrow}({\mathbf r})\}\rangle\\
		\langle\{\delta\hat\psi^\dagger_{\downarrow}({\mathbf r}'), \delta\hat\psi_{\uparrow}({\mathbf r})\}\rangle & \langle\{\delta\hat\psi^\dagger_{\downarrow}({\mathbf r}'), \delta\hat\psi_{\downarrow}({\mathbf r})\}\rangle
	\end{pmatrix}
\end{align}
and 
\begin{align}
	\Gamma_{12}( \mathbf{r}, \mathbf{r}^{\prime} ) = \begin{pmatrix}
		\langle\{\delta\hat\psi_{\uparrow}({\mathbf r}'), \delta\hat\psi_{\uparrow}({\mathbf r})\}\rangle & \langle\{\delta\hat\psi_{\uparrow}({\mathbf r}'), \delta\hat\psi_{\downarrow}({\mathbf r})\}\rangle\\
		\langle\{\delta\hat\psi_{\downarrow}({\mathbf r}'), \delta\hat\psi_{\uparrow}({\mathbf r})\}\rangle & \langle\{\delta\hat\psi_{\downarrow}({\mathbf r}'), \delta\hat\psi_{\downarrow}({\mathbf r})\}\rangle
	\end{pmatrix}
\end{align}
and $\Gamma_{21} = \Gamma_{12}^{\dagger}$, $\Gamma_{22} = \Gamma_{11}^{T}$. It can be shown that $\Gamma$ and $\xi$ are connected through $\Gamma = S S^{\dagger}$, where $S = e^{i \Sigma^{z} \xi}$ is a symplectic matrix satisfying $S \Sigma^{z} S^{\dagger} = \Sigma^{z}$. So it is convenient to take the elements of $\phi^{(c)}({\mathbf r})$ and $\Gamma( \mathbf{r}, \mathbf{r}^{\prime} )$ as variational parameters.

To proceed further, we substitute the expansion $\psi( \mathbf{r} ) = \phi^{(c)}( \mathbf{r} ) + \delta \psi( \mathbf{r} ) $ into the Hamiltonian $H$ defined in main text, which leads to single-particle Hamiltonian
\begin{align}\label{1}
	H_{0} = \sum_{\alpha = \uparrow, \downarrow} \int d \mathbf{r} \left( \phi^{(c)}_{\alpha}( \mathbf{r} )^{\dagger} h_{\alpha} \phi^{(c)}_{\alpha}( \mathbf{r} ) + \delta \hat{\psi}^{\dagger}_{\alpha}( \mathbf{r} ) h_{\alpha} \phi^{(c)}( \mathbf{r} ) + \phi^{(c)}_{\alpha}( \mathbf{r} )^{\dagger} h_{\alpha} \delta \hat{\psi}_{\alpha}( \mathbf{r} ) + \delta \hat{\psi}^{\dagger}_{\alpha}( \mathbf{r} ) h_{\alpha} \delta \hat{\psi}_{\alpha}( \mathbf{r} ) \right)
\end{align}
two-body interaction Hamiltonian
\begin{align}
	H_{2B} &= \sum_{\alpha,\beta=\uparrow,\downarrow} \frac{g_{\alpha \beta}}{2} \int d \mathbf{r} \left[ |\phi^{(c)}_{\alpha}(\mathbf{r})|^{2} |\phi^{(c)}_{\beta}(\mathbf{r})|^{2} + \left( 2 \delta \hat{\psi}^{\dagger}_{\alpha}(\mathbf{r}) \phi^{(c)}_{\alpha}(\mathbf{r}) |\phi^{(c)}_{\beta}(\mathbf{r})|^{2} + \delta \hat{\psi}^{\dagger}_{\alpha}(\mathbf{r}) \delta \hat{\psi}^{\dagger}_{\beta}(\mathbf{r}) \phi^{(c)}_{\beta}(\mathbf{r}) \phi^{(c)}_{\alpha}(\mathbf{r}) \right. \right. \notag \\
	&\quad+ 2 \delta \hat{\psi}^{\dagger}_{\alpha}(\mathbf{r}) \delta \hat{\psi}^{\dagger}_{\beta}(\mathbf{r}) \delta \hat{\psi}_{\beta}(\mathbf{r}) \phi^{(c)}_{\alpha}(\mathbf{r}) + H.c. \Big) + 2 \delta \hat{\psi}^{\dagger}_{\alpha}(\mathbf{r})  \delta \hat{\psi}_{\alpha}(\mathbf{r}) |\phi^{(c)}_{\beta}(\mathbf{r})|^{2} + 2 \delta \hat{\psi}^{\dagger}_{\alpha}(\mathbf{r})  \delta \hat{\psi}_{\beta}(\mathbf{r}) \phi^{(c) \ast}_{\beta}(\mathbf{r}) \phi^{(c)}_{\alpha}(\mathbf{r}) \notag \\
	&\quad+ \delta \hat{\psi}^{\dagger}_{\alpha}(\mathbf{r}) \delta \hat{\psi}^{\dagger}_{\beta}(\mathbf{r}) \delta \hat{\psi}_{\beta}(\mathbf{r}) \delta \hat{\psi}_{\alpha}(\mathbf{r}) \Big]
\end{align}
and three-body interaction Hamiltonian $H_{3B}$
\begin{align}
	H_{3B} &= \frac{g_{3}}{3!} \sum_{\alpha, \beta, \gamma = \uparrow, \downarrow} \int d\mathbf{r} \left[ |\phi^{(c)}_{\alpha}(\mathbf{r})|^{2} |\phi^{(c)}_{\beta}(\mathbf{r})|^{2} |\phi^{(c)}_{\gamma}(\mathbf{r})|^{2} + \left( 3 \delta \hat{\psi}^{\dagger}_{\alpha}(\mathbf{r}) \phi^{(c)}_{\alpha}(\mathbf{r}) |\phi^{(c)}_{\beta}(\mathbf{r})|^{2} |\phi^{(c)}_{\gamma}(\mathbf{r})|^{2} \right. \right. \notag \\
	&\quad+ 3 \delta \hat{\psi}^{\dagger}_{\alpha}(\mathbf{r}) \delta \hat{\psi}^{\dagger}_{\beta}(\mathbf{r}) \phi^{(c)}_{\beta}(\mathbf{r}) \phi^{(c)}_{\alpha}(\mathbf{r}) |\phi^{(c)}_{\gamma}(\mathbf{r})|^{2} + 3 \delta \hat{\psi}^{\dagger}_{\alpha}(\mathbf{r}) \delta \hat{\psi}_{\beta}(\mathbf{r}) \phi^{(c) \ast}_{\beta}(\mathbf{r}) \phi^{(c)}_{\alpha}(\mathbf{r}) |\phi^{(c)}_{\gamma}(\mathbf{r})|^{2} \notag \\
	&\quad+ \delta \hat{\psi}^{\dagger}_{\alpha}(\mathbf{r}) \delta \hat{\psi}^{\dagger}_{\beta}(\mathbf{r}) \delta \hat{\psi}^{\dagger}_{\gamma}(\mathbf{r}) \phi^{(c)}_{\gamma}(\mathbf{r}) \phi^{(c)}_{\beta}(\mathbf{r}) \phi^{(c)}_{\alpha}(\mathbf{r}) + 6 \delta \hat{\psi}^{\dagger}_{\alpha}(\mathbf{r}) \delta \hat{\psi}^{\dagger}_{\beta}(\mathbf{r}) \delta \hat{\psi}_{\alpha}(\mathbf{r}) \phi^{(c)}_{\beta}(\mathbf{r}) |\phi^{(c)}_{\gamma}(\mathbf{r})|^{2} \notag \\
	&\quad+ 3 \delta \hat{\psi}^{\dagger}_{\alpha}(\mathbf{r}) \delta \hat{\psi}^{\dagger}_{\beta}(\mathbf{r}) \delta \hat{\psi}_{\gamma}(\mathbf{r}) \phi^{(c) \ast}_{\gamma}(\mathbf{r}) \phi^{(c)}_{\beta}(\mathbf{r}) \phi^{(c)}_{\alpha}(\mathbf{r}) + 3 \delta \hat{\psi}^{\dagger}_{\alpha}(\mathbf{r}) \delta \hat{\psi}^{\dagger}_{\beta}(\mathbf{r}) \delta \hat{\psi}^{\dagger}_{\gamma}(\mathbf{r}) \delta \hat{\psi}_{\gamma}(\mathbf{r}) \phi^{(c)}_{\beta}(\mathbf{r}) \phi^{(c)}_{\alpha}(\mathbf{r}) \notag \\
	&\quad+ 3 \delta \hat{\psi}^{\dagger}_{\alpha}(\mathbf{r}) \delta \hat{\psi}^{\dagger}_{\beta}(\mathbf{r}) \delta \hat{\psi}^{\dagger}_{\gamma}(\mathbf{r}) \delta \hat{\psi}_{\gamma}(\mathbf{r}) \delta \hat{\psi}_{\beta}(\mathbf{r}) \phi^{(c)}_{\alpha}(\mathbf{r}) + H.c. \Big) + 3 \delta \hat{\psi}^{\dagger}_{\alpha}(\mathbf{r}) \delta \hat{\psi}_{\alpha}(\mathbf{r}) |\phi^{(c)}_{\beta}(\mathbf{r})|^{2} |\phi^{(c)}_{\gamma}(\mathbf{r})|^{2} \notag \\
	&\quad+ 3 \delta \hat{\psi}^{\dagger}_{\alpha}(\mathbf{r}) \delta \hat{\psi}^{\dagger}_{\beta}(\mathbf{r}) \delta \hat{\psi}_{\beta}(\mathbf{r}) \delta \hat{\psi}_{\alpha}(\mathbf{r}) |\phi^{(c)}_{\gamma}(\mathbf{r})|^{2} + 6 \delta \hat{\psi}^{\dagger}_{\alpha}(\mathbf{r}) \delta \hat{\psi}^{\dagger}_{\beta}(\mathbf{r}) \delta \hat{\psi}_{\gamma}(\mathbf{r}) \delta \hat{\psi}_{\beta}(\mathbf{r}) \phi^{(c) \ast}_{\gamma}(\mathbf{r}) \phi^{(c)}_{\alpha}(\mathbf{r}) \notag \\
	&\quad+ \delta \hat{\psi}^{\dagger}_{\alpha}(\mathbf{r}) \delta \hat{\psi}^{\dagger}_{\beta}(\mathbf{r}) \delta \hat{\psi}^{\dagger}_{\gamma}(\mathbf{r}) \delta \hat{\psi}_{\gamma}(\mathbf{r}) \delta \hat{\psi}_{\beta}(\mathbf{r}) \delta \hat{\psi}_{\alpha}(\mathbf{r}) \Big]
\end{align}
After applying the Wick's theorem with respect to the Gaussian state $| \Psi_{\rm{GS}} \rangle,$ we obtain the mean-field Hamiltonian
\begin{align}
	H_{\rm{MF}} = E_0 + (\delta \hat{\psi}^{\dagger} \eta + \eta^{\ast} \delta \hat{\psi}) + \frac{1}{2} : \delta \hat{\Psi}^{\dagger} \mathcal{H} \delta \hat{\Psi}:,
\end{align}
where $\mathopen{:}\hat{O}\mathclose{:}$ denotes the normal-ordered operator defined with respect to the Gaussian state, $E_{0} \equiv \left\langle \Psi_{\rm GS} \right| H \left| \Psi_{\rm GS} \right\rangle$ is the expectation value of the Hamiltonian and is composed of the kinetic energy 
\begin{align}
	E_{k} = \sum_{\alpha = \uparrow, \downarrow} \int d\mathbf{r} \left( \phi^{(c) \ast}_{\alpha}(\mathbf{r}) h_{\alpha} \phi^{(c)}_{\alpha}(\mathbf{r}) + \langle \delta\hat{\psi}^{\dagger}_{\alpha}(\mathbf{r}) h_{\alpha} \delta\hat{\psi}_{\alpha}(\mathbf{r}) \rangle \right),
\end{align}
the two-body interaction energy
\begin{align}
	E_{2B} &= \sum_{\alpha, \beta = \uparrow, \downarrow} \frac{g_{\alpha\beta}}{2} \int d\mathbf{r} \left[ |\phi^{(c)}_{\alpha}(\mathbf{r})|^{2} |\phi^{(c)}_{\beta}(\mathbf{r})|^{2} + 2 \rm{Re} \left( \langle \delta\hat{\psi}_{\alpha}(\mathbf{r}) \delta\hat{\psi}_{\beta}(\mathbf{r}) \rangle \phi^{(c) \ast}_{\beta}(\mathbf{r}) \phi^{(c) \ast}_{\alpha}(\mathbf{r}) \right) + |\langle \delta\hat{\psi}_{\alpha}(\mathbf{r}) \delta\hat{\psi}_{\beta}(\mathbf{r}) \rangle|^{2} \right. \notag \\
	&\quad \left. + \left( 2 |\phi^{(c)}_{\alpha}(\mathbf{r})|^{2} + \langle \delta\hat{\psi}^{\dagger}_{\alpha}(\mathbf{r}) \delta\hat{\psi}_{\alpha}(\mathbf{r}) \rangle \right) \langle \delta\hat{\psi}^{\dagger}_{\beta}(\mathbf{r}) \delta\hat{\psi}_{\beta}(\mathbf{r}) \rangle + \left( 2 \phi^{(c) \ast}_{\alpha}(\mathbf{r}) \phi^{(c)}_{\beta}(\mathbf{}r) + \langle \delta\hat{\psi}^{\dagger}_{\alpha}(\mathbf{r}) \delta\hat{\psi}_{\beta}(\mathbf{r}) \rangle \right) \langle \delta\hat{\psi}^{\dagger}_{\beta}(\mathbf{r}) \delta\hat{\psi}_{\alpha}(\mathbf{r}) \rangle \right],
\end{align}
and the tree-body interaction energy
\begin{align}
	E_{3B} &= \frac{g_{3}}{3!} \sum_{\alpha, \beta, \gamma = \uparrow, \downarrow} \int d\mathbf{r} \left[ |\phi^{(c)}_{\alpha}(\mathbf{r})|^{2} |\phi^{(c)}_{\beta}(\mathbf{r})|^{2} |\phi^{(c)}_{\gamma}(\mathbf{r})|^{2} + 6\rm{Re}\left( \langle \delta \hat{\psi}^{\dagger}_{\alpha}(\mathbf{r}) \delta \hat{\psi}^{\dagger}_{\beta}(\mathbf{r}) \rangle \phi^{(c)}_{\beta}(\mathbf{r}) \phi^{(c)}_{\alpha}(\mathbf{r}) |\phi^{(c)}_{\gamma}(\mathbf{r})|^{2} \right) \right. \notag \\
	&\quad+ 6 \langle \delta \hat{\psi}^{\dagger}_{\alpha}(\mathbf{r}) \delta \hat{\psi}_{\beta}(\mathbf{r}) \rangle \phi^{(c) \ast}_{\beta}(\mathbf{r}) \phi^{(c)}_{\alpha}(\mathbf{r}) |\phi^{(c)}_{\gamma}(\mathbf{r})|^{2} + 3 \langle \delta \hat{\psi}^{\dagger}_{\alpha}(\mathbf{r}) \delta \hat{\psi}_{\alpha}(\mathbf{r}) \rangle |\phi^{(c)}_{\beta}(\mathbf{r})|^{2} |\phi^{(c)}_{\gamma}(\mathbf{r})|^{2} \notag \\
	&\quad+ 2 \langle \delta\hat{\psi}_{\alpha}(\mathbf{r}) \delta\hat{\psi}_{\beta}(\mathbf{r}) \rangle \left( \langle \delta \hat{\psi}^{\dagger}_{\alpha}(\mathbf{r}) \delta \hat{\psi}^{\dagger}_{\beta}(\mathbf{r}) \rangle \langle \delta \hat{\psi}^{\dagger}_{\gamma}(\mathbf{r}) \delta \hat{\psi}_{\gamma}(\mathbf{r}) \rangle + 2 \langle \delta \hat{\psi}^{\dagger}_{\alpha}(\mathbf{r}) \delta \hat{\psi}^{\dagger}_{\gamma}(\mathbf{r}) \rangle \langle \delta \hat{\psi}^{\dagger}_{\beta}(\mathbf{r}) \delta \hat{\psi}_{\gamma}(\mathbf{r}) \rangle \right) \notag \\
	&\quad+ 6\rm{Re}\left( \phi^{(c)}_{\beta}(\mathbf{r}) \phi^{(c)}_{\alpha}(\mathbf{r}) \left( \langle \delta \hat{\psi}^{\dagger}_{\alpha}(\mathbf{r}) \delta \hat{\psi}^{\dagger}_{\beta}(\mathbf{r}) \rangle \langle \delta \hat{\psi}^{\dagger}_{\gamma}(\mathbf{r}) \delta \hat{\psi}_{\gamma}(\mathbf{r}) \rangle + 2 \langle \delta \hat{\psi}^{\dagger}_{\alpha}(\mathbf{r}) \delta \hat{\psi}^{\dagger}_{\gamma}(\mathbf{r}) \rangle \langle \delta \hat{\psi}^{\dagger}_{\beta}(\mathbf{r}) \delta \hat{\psi}_{\gamma}(\mathbf{r}) \rangle \right) \right) \notag \\
	&\quad+ \left( 3 |\phi^{(c)}_{\alpha}(\mathbf{r})|^{2} + \langle \delta \hat{\psi}^{\dagger}_{\alpha}(\mathbf{r}) \delta \hat{\psi}_{\alpha}(\mathbf{r}) \rangle \right) \left( |\langle \delta \hat{\psi}_{\beta}(\mathbf{r}) \delta \hat{\psi}_{\gamma}(\mathbf{r}) \rangle|^{2} + \langle \delta \hat{\psi}^{\dagger}_{\gamma}(\mathbf{r}) \delta \hat{\psi}_{\gamma}(\mathbf{r}) \rangle \langle \delta \hat{\psi}^{\dagger}_{\beta}(\mathbf{r}) \delta \hat{\psi}_{\beta}(\mathbf{r}) \rangle \right. \notag\\
	&\quad+ |\langle \delta \hat{\psi}^{\dagger}_{\gamma}(\mathbf{r}) \delta \hat{\psi}_{\beta}(\mathbf{r}) \rangle|^{2} \Big) + 2 \left( 3 \phi^{(c) \ast}_{\beta}(\mathbf{r}) \phi^{(c)}_{\alpha}(\mathbf{r}) + \langle \delta \hat{\psi}^{\dagger}_{\beta}(\mathbf{r}) \delta \hat{\psi}_{\alpha}(\mathbf{r}) \rangle \right) \left( \langle \delta \hat{\psi}^{\dagger}_{\alpha}(\mathbf{r}) \delta \hat{\psi}^{\dagger}_{\gamma}(\mathbf{r}) \rangle \langle \delta \hat{\psi}_{\gamma}(\mathbf{r}) \delta \hat{\psi}_{\beta}(\mathbf{r}) \rangle \right. \notag \\
	&\quad+ \langle \delta \hat{\psi}^{\dagger}_{\alpha}(\mathbf{r}) \delta \hat{\psi}_{\gamma}(\mathbf{r}) \rangle \langle \delta \hat{\psi}^{\dagger}_{\gamma}(\mathbf{r}) \delta \hat{\psi}_{\beta}(\mathbf{r}) \rangle + \langle \delta \hat{\psi}^{\dagger}_{\alpha}(\mathbf{r}) \delta \hat{\psi}_{\beta}(\mathbf{r}) \rangle \langle \delta \hat{\psi}^{\dagger}_{\gamma}(\mathbf{r}) \delta \hat{\psi}_{\gamma}(\mathbf{r}) \rangle \Big) \Big].
\end{align}
The linear driving term $\eta = \begin{pmatrix}
	\eta_{\uparrow} \\
	\eta_{\downarrow}
\end{pmatrix} $ is composed of
\begin{align}
	\eta_{\alpha} &= h_{\alpha} \phi^{(c)}_{\alpha} + \sum_{\beta = \uparrow \downarrow} g_{\alpha\beta} \left( \left( |\phi^{(c)}_{\beta}|^{2} + \langle \delta\hat{\psi}^{\dagger}_{\beta} \delta\hat{\psi}_{\beta} \rangle \right) \phi^{(c)}_{\alpha} + \langle \delta\hat{\psi}_{\alpha} \delta\hat{\psi}_{\beta} \rangle \phi^{(c) \ast}_{\beta} + \langle \delta\hat{\psi}^{\dagger}_{\beta} \delta\hat{\psi}_{\alpha} \rangle \phi^{(c)}_{\beta} \right) \notag \\
	&\quad+ \frac{g_{3}}{2} \sum_{\beta, \gamma = \uparrow, \downarrow} \left[ \left( |\phi^{(c)}_{\beta}|^{2} |\phi^{(c)}_{\gamma}|^{2} + 2\rm{Re}\left( \langle \delta \hat{\psi}^{\dagger}_{\beta} \delta \hat{\psi}^{\dagger}_{\gamma} \rangle \phi^{(c)}_{\gamma} \phi^{(c)}_{\beta} \right) + 2 \langle \delta \hat{\psi}^{\dagger}_{\gamma} \delta \hat{\psi}_{\beta} \rangle \phi^{(c) \ast}_{\beta} \phi^{(c)}_{\gamma} + 2 \langle \delta \hat{\psi}^{\dagger}_{\beta} \delta \hat{\psi}_{\beta} \rangle |\phi^{(c)}_{\gamma}|^{2} \right. \right. \notag \\
	&\quad+ |\langle \delta \hat{\psi}_{\beta} \delta \hat{\psi}_{\gamma} \rangle|^{2} + \langle \delta \hat{\psi}^{\dagger}_{\gamma} \delta \hat{\psi}_{\gamma} \rangle \langle \delta \hat{\psi}^{\dagger}_{\beta} \delta \hat{\psi}_{\beta} \rangle + |\langle \delta \hat{\psi}^{\dagger}_{\gamma} \delta \hat{\psi}_{\beta} \rangle|^{2} \Big) \phi^{(c)}_{\alpha} + 2 \left( \langle \delta \hat{\psi}_{\alpha} \delta \hat{\psi}_{\beta} \rangle |\phi^{(c)}_{\gamma}|^{2} + \langle \delta\hat{\psi}_{\alpha} \delta\hat{\psi}_{\beta} \rangle \langle \delta\hat{\psi}^{\dagger}_{\gamma} \delta\hat{\psi}_{\gamma} \rangle \right. \notag \\
	&\quad+ \langle \delta\hat{\psi}^{\dagger}_{\gamma} \delta\hat{\psi}_{\beta} \rangle \langle \delta\hat{\psi}_{\gamma} \delta\hat{\psi}_{\alpha} \rangle + \langle \delta\hat{\psi}^{\dagger}_{\gamma} \delta\hat{\psi}_{\alpha} \rangle \langle \delta\hat{\psi}_{\gamma} \delta\hat{\psi}_{\beta} \rangle \Big) \phi^{(c) \ast}_{\beta} + 2 \left( \langle \delta \hat{\psi}^{\dagger}_{\beta} \delta \hat{\psi}_{\alpha} \rangle |\phi^{(c)}_{\gamma}|^{2} + \langle \delta \hat{\psi}^{\dagger}_{\beta} \delta \hat{\psi}^{\dagger}_{\gamma} \rangle \langle \delta \hat{\psi}_{\gamma} \delta \hat{\psi}_{\alpha} \rangle \right. \notag \\
	&\quad+ \langle \delta \hat{\psi}^{\dagger}_{\beta} \delta \hat{\psi}_{\gamma} \rangle \langle \delta \hat{\psi}^{\dagger}_{\gamma} \delta \hat{\psi}_{\alpha} \rangle + \langle \delta \hat{\psi}^{\dagger}_{\beta} \delta \hat{\psi}_{\alpha} \rangle \langle \delta \hat{\psi}^{\dagger}_{\gamma} \delta \hat{\psi}_{\gamma} \rangle \Big) \phi^{(c)}_{\beta} \Big].\label{eta}
\end{align}
The matrix elements of $\mathcal{H} = \begin{pmatrix}
	\mathcal{E} & \Delta \\
	\Delta^{\dagger} & \mathcal{E}^{\ast}
\end{pmatrix} $ are defined as $\mathcal{E} = \begin{pmatrix}
	\mathcal{E}_{\uparrow \uparrow} & \mathcal{E}_{\uparrow \downarrow} \\
	\mathcal{E}_{\downarrow \uparrow} & \mathcal{E}_{\downarrow \downarrow}
\end{pmatrix} $ and $\Delta = \begin{pmatrix}
	\Delta_{\uparrow \uparrow} & \Delta_{\uparrow \downarrow} \\
	\Delta_{\downarrow \uparrow} & \Delta_{\downarrow \downarrow}
\end{pmatrix} $, where
\begin{align}
	\mathcal{E}_{\alpha \beta} &= \sum_{\gamma = \uparrow, \downarrow} \left[ h_{\alpha} + g_{\alpha \gamma} \left( |\phi^{(c)}_{\gamma}|^{2} + \langle \delta\hat{\psi}^{\dagger}_{\gamma} \delta\hat{\psi}_{\gamma} \rangle \right) \right] \delta_{\alpha \beta} + g_{\alpha \beta} \left( \phi^{(c)}_{\alpha} \phi^{(c) \ast}_{\beta} + \langle \delta \hat{\psi}^{\dagger}_{\beta} \delta\hat{\psi}_{\alpha} \rangle \right) \notag \\
	&\quad+ \frac{g_{3}}{3!} \sum_{\gamma = \uparrow, \downarrow} \left[ \sum_{\gamma^{\prime} = \uparrow, \downarrow} \left( \frac{1}{2} |\phi^{(c)}_{\gamma}|^{2} |\phi^{(c)}_{\gamma^{\prime}}|^{2} + |\phi^{(c)}_{\gamma}|^{2} \langle \delta\hat{\psi}^{\dagger}_{\gamma^{\prime}} \delta\hat{\psi}_{\gamma^{\prime}} \rangle + \phi^{(c) \ast}_{\gamma^{\prime}} \phi^{(c)}_{\gamma} \langle \delta \hat{\psi}^{\dagger}_{\gamma} \delta\hat{\psi}_{\gamma^{\prime}} \rangle \right. \right. \notag \\
	&\quad+ \rm{Re}\left( \phi^{(c) \ast}_{\gamma} \phi^{(c) \ast}_{\gamma^{\prime}} \langle \delta \hat{\psi}_{\gamma^{\prime}} \delta \hat{\psi}_{\gamma} \rangle \right) + \frac{1}{2} \left( \langle \delta\hat{\psi}^{\dagger}_{\gamma} \delta\hat{\psi}_{\gamma} \rangle \langle \delta\hat{\psi}^{\dagger}_{\gamma^{\prime}} \delta\hat{\psi}_{\gamma^{\prime}} \rangle + \langle \delta\hat{\psi}^{\dagger}_{\gamma} \delta\hat{\psi}_{\gamma^{\prime}} \rangle \langle \delta\hat{\psi}^{\dagger}_{\gamma^{\prime}} \delta\hat{\psi}_{\gamma} \rangle \right. \notag \\
	&\quad+ \langle \delta \hat{\psi}^{\dagger}_{\gamma} \delta \hat{\psi}^{\dagger}_{\gamma^{\prime}} \rangle \langle \delta \hat{\psi}_{\gamma^{\prime}} \delta \hat{\psi}_{\gamma} \rangle \Big) \Big) \delta_{\alpha \beta} + \left( |\phi^{(c)}_{\gamma}|^{2} + \langle \delta\hat{\psi}^{\dagger}_{\gamma} \delta\hat{\psi}_{\gamma} \rangle \right) \left( \phi^{(c)}_{\alpha} \phi^{(c) \ast}_{\beta} + \langle \delta \hat{\psi}^{\dagger}_{\beta} \delta\hat{\psi}_{\alpha} \rangle \right) \notag \\
	&\quad+ \phi^{(c)}_{\gamma} \phi^{(c) \ast}_{\beta} \langle \delta \hat{\psi}^{\dagger}_{\gamma} \delta\hat{\psi}_{\alpha} \rangle + \phi^{(c)}_{\alpha} \phi^{(c) \ast}_{\gamma} \langle \delta \hat{\psi}^{\dagger}_{\beta} \delta\hat{\psi}_{\gamma} \rangle + \phi^{(c)}_{\alpha} \phi^{(c)}_{\gamma} \langle \delta\hat{\psi}^{\dagger}_{\gamma} \delta \hat{\psi}^{\dagger}_{\beta} \rangle + \phi^{(c) \ast}_{\gamma} \phi^{(c) \ast}_{\beta} \langle \delta \hat{\psi}_{\gamma} \delta\hat{\psi}_{\alpha} \rangle \notag \\
	&\quad+ \langle \delta \hat{\psi}^{\dagger}_{\gamma} \delta\hat{\psi}_{\alpha} \rangle \langle \delta \hat{\psi}^{\dagger}_{\beta} \delta\hat{\psi}_{\gamma} \rangle + \langle \delta\hat{\psi}_{\alpha} \delta \hat{\psi}_{\gamma} \rangle \langle \delta\hat{\psi}^{\dagger}_{\gamma} \delta \hat{\psi}^{\dagger}_{\beta} \rangle \Bigg]\label{cale}
\end{align}
and
\begin{align}
	\Delta_{\alpha \beta} &= g_{\alpha \beta} \left( \phi^{(c)}_{\alpha} \phi^{(c)}_{\beta} + \langle \delta\hat{\psi}_{\alpha} \delta \hat{\psi}_{\beta} \rangle \right) + g_{3} \sum_{\gamma = \uparrow, \downarrow} \left[ \left( |\phi^{(c)}_{\gamma}|^{2} + \langle \delta\hat{\psi}^{\dagger}_{\gamma} \delta\hat{\psi}_{\gamma} \rangle \right) \left( \phi^{(c)}_{\alpha} \phi^{(c)}_{\beta} + \langle \delta\hat{\psi}_{\alpha} \delta \hat{\psi}_{\beta} \rangle \right) \right. \notag \\
	&\quad+ \phi^{(c)}_{\beta} \phi^{(c)}_{\gamma} \langle \delta\hat{\psi}^{\dagger}_{\gamma} \delta\hat{\psi}_{\alpha} \rangle + \phi^{(c)}_{\alpha} \phi^{(c)}_{\gamma} \langle \delta\hat{\psi}^{\dagger}_{\gamma} \delta\hat{\psi}_{\beta} \rangle + \langle \delta\hat{\psi}_{\alpha} \delta \hat{\psi}_{\gamma} \rangle \phi^{(c) \ast}_{\gamma} \phi^{(c)}_{\beta} + \langle \delta\hat{\psi}_{\beta} \delta \hat{\psi}_{\gamma} \rangle \phi^{(c) \ast}_{\gamma} \phi^{(c)}_{\alpha} \notag \\
	&\quad+ \langle \delta\hat{\psi}_{\alpha} \delta \hat{\psi}_{\gamma} \rangle \langle \delta\hat{\psi}^{\dagger}_{\gamma} \delta\hat{\psi}_{\beta} \rangle + \langle \delta\hat{\psi}_{\beta} \delta \hat{\psi}_{\gamma} \rangle \langle \delta\hat{\psi}^{\dagger}_{\gamma} \delta\hat{\psi}_{\alpha} \rangle \Big].\label{delta}
\end{align}
Based on the GST, the ground-state solutions, $\phi^{(c)}({\mathbf r})$ and $\Gamma({\mathbf r},{\mathbf r}')$, can be obtained by numerically evolving the imaginary-time equations of motion~\cite{Shi2018_SM,Shi2019_SM},
\begin{subequations}
	\label{EOM}
	\begin{align}
		\partial _{\tau }\Phi & =-\Gamma
		\begin{pmatrix}
			\eta \\
			\eta ^{\ast }%
		\end{pmatrix}%
		,  \label{Imva} \\
		\partial _{\tau }\Gamma & =\Sigma ^{z}\mathcal{H}\Sigma ^{z}-\Gamma \mathcal{%
			H}\Gamma,  \label{Imvb}
	\end{align}
\end{subequations}
which converge when the imaginary time $\tau$ is sufficiently large.

\section{Atom-number distribution}\label{secand}

\begin{figure}[ptb]
\centering
\includegraphics[width=0.45\columnwidth]{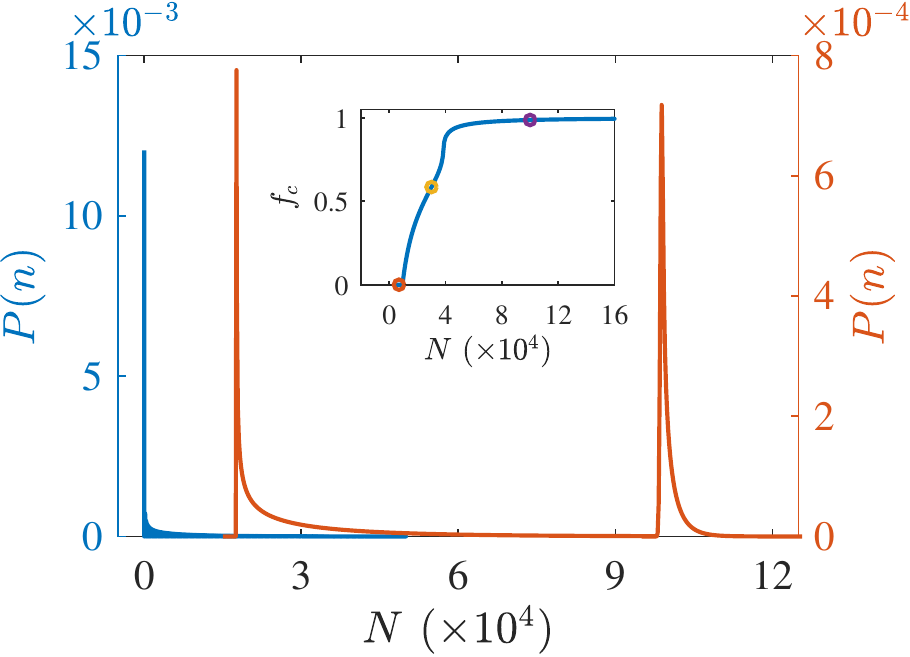}
\caption{(color online). Atom number distributions $P(n)$ for the states with mean atom numbers (from left to right) $N=6.9\times 10^3$, $3\times 10^4$, and $10^5$. The $y$ axis of the blue line (red lines) is on the left (right). The inset shows the coherent fraction as a function of mean atom number where the markers denote the mean atom numbers used to plot $P(n)$'s. The reduced scattering length is $\delta a=-3.062 a_{0}$.}
\label{pnd}
\end{figure}

Following the approach used in dipolar droplets~\cite{Wang2020_SM}, here we derive the atom-number distribution of a Gaussian state describing a binary condensate. To this end, we expand the coherent mode $\hat{c}$ in terms of the squeezed modes as
\begin{align}
	\hat{c}=\sum_{i=1}^{\infty} \alpha_{i} \hat{s}_{i}+\alpha_{\perp} \hat{c}_{\perp}
\end{align}
where $\alpha_{i}=\int d \mathbf{r} \left[\bar{\phi}_{i}^{(s)}(\mathbf{r})\right]^\dag \bar{\phi}^{(c)}(\mathbf{r})$, $\alpha_{\perp}^{2}=1-\sum_{i}\left|\alpha_{i}\right|^{2}$, and $\hat{c}_{\perp}$ represents the mode that is perpendicular to all $\hat{s}_{i}$. The Gaussian state wave function can now be expressed as
\begin{align}
	\left| \Psi_{\mathrm{GS}} \right\rangle
	&= e^{\sqrt{N^{(c)}} \alpha_{\perp} \left( \hat{c}_{\perp}^{\dagger} - \hat{c}_{\perp} \right)}\prod^{\infty}_{i=1} e^{\sqrt{N^{(c)}} \left( \alpha_{i} \hat{s}_{i}^{\dagger} - \alpha_{i}^{*} \hat{s}_{i} \right)} e^{\frac{1}{2} \xi_{i}\left(\hat{s}^{\dagger 2}_{i}-\hat{s}^{2}_{i} \right)} \left| 0 \right\rangle.
\end{align}
The AND for mode $\hat{c}_{\perp}$ is $$p_{\perp}(\ell)=e^{-N^{(c)} \left|\alpha_{\perp}\right|^{2}} \left( N^{(c)}\left|\alpha_{\perp}\right|^{2} \right)^{\ell}/ \ell !$$ and that for mode $\hat{s}_{i}$ is \cite{Knight2005_SM}
\begin{align}
	p_{i}(\ell)=& \frac{\left(\frac{1}{2} \tanh \xi_{i}\right)^{\ell}}{\ell ! \cosh \xi_{i}}\left|H_{\ell}\left[\gamma_{i}\left(\sinh 2 \xi_{i}\right)^{-1 / 2}\right]\right|^{2} \exp \left[-N^{(c)}\left|\alpha_{i}\right|^{2}-\frac{N^{(c)}}{2}\left(\alpha_{i}^{* 2}+\alpha_{i}^{2}\right) \tanh \xi_{i}\right],
\end{align}
where $\gamma_{i}=\sqrt{N^{(c)}} \alpha_{i} \cosh \xi_{i}+\sqrt{N^{(c)}} \alpha_{i}^{*} \sinh \xi_{i}$ and $H_{\ell}(x)$ are Hermite polynomials. The AND of $\left|\Psi_{\mathrm{GS}}\right\rangle$ can then be expressed into a recursive form as follows. Let $P_{i}(n)$ denote the AND of the state containing first $i$ squeezed modes, the AND containing first $i+1$ squeezed modes can then be expressed as
\begin{align}
	P_{i+1}(n)=\sum_{\ell=0}^{n} P_{i}(n-\ell) p_{i+1}(\ell)
\end{align}
where $P_{0}(n) \equiv p_{\perp}(n)$. Applying this equation successively will eventually leads to the total AND $P(n)$.

Figure~\ref{pnd} shows three typical ANDs corresponding to states in SVS, SCS, and CS phases. Due to the squeezing, $P(n)$ is asymmetric with a long tail at the large $n$ side, which was observed in dipolar droplet experiments~\cite{asypnd-1,asypnd-2}. However, as $f_c$ increases, $P(n)$ becomes more symmetric. Other features of the AND include that $P(n)$ has a peak at $n\approx N^{(c)}$ and the mean value of atom number is $$\sum_nnP(n)=N^{(c)}+N^{(s)}=N.$$

\section{Virial Relation}\label{secvirial}

\begin{figure}[ptb]
\centering
\includegraphics[width=0.45\columnwidth]{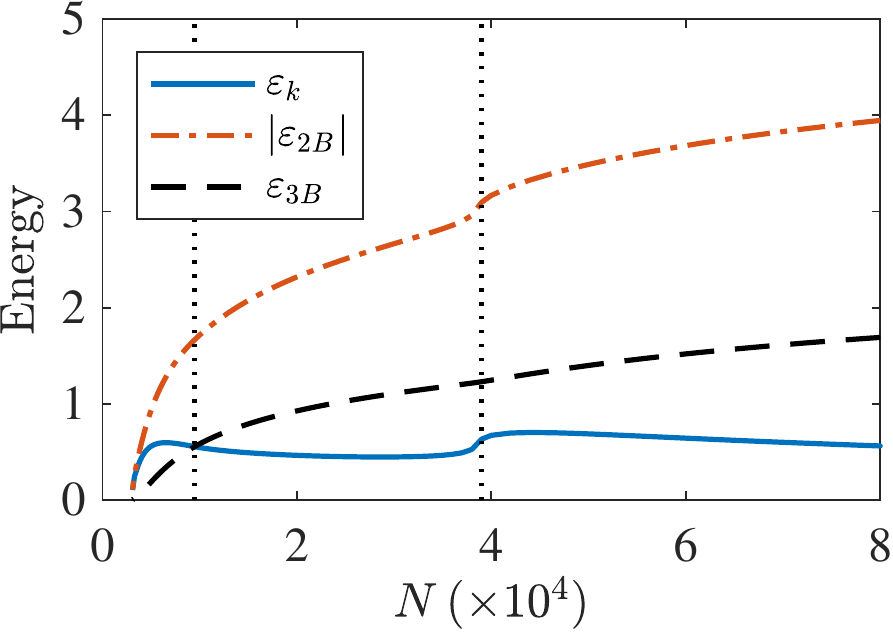}
\caption{(color online). $\varepsilon_{k}$, $\varepsilon_{2B}$, and $\varepsilon_{3B}$ as functions of $N$ for $\delta a=-3.062 a_{0}$. Energy is in units of $\hbar^2/(ML^2)$ with $L=6\,\mu{\rm m}$. Vertical dotted lines mark the locations of two phase transitions.}
	\label{energy}
\end{figure}

Here we derive a virial relation among different contributions to the total energy~\cite{Stringari_SM}. To this end, we shall make use of the fact that different spin component share same normalized density profile. Hence we can use $\bar{\phi}^{(c)}(\mathbf{r})$ to represent the normalized mode function for the coherent component and for $i$th squeezed mode the normalized mode functions are $\bar{\phi}^{(s)}_{i}(\mathbf{r})$. If total energy $E_0 = E_{k} + E_{2B} + E_{3B}$ is stationary for any variation of $\bar{\phi}^{(c)}(\mathbf{r})$ and $\bar{\phi}^{(s)}_{i}(\mathbf{r})$ around the ground state solution, we can choose scaling transformations of the form $\phi(x, y) \rightarrow (1 + \lambda)^{1/2} \phi[(1 + \lambda) x, y]$ and insert kinetic energy, 2B and 3B interaction energies, we have
\begin{align}
	\delta E_{k} =& [(1 + \lambda)^{2} - 1] E^{x}_{k} \notag \\
	\delta E_{2B} =& \lambda E_{2B} \notag \\
	\delta E_{3B} =& [(1 + \lambda)^{2} - 1] E_{3B},
\end{align}
where
\begin{align}
	E^{x}_{k} = - \frac{1}{2 M} \int d\mathbf{r} \left( N^{(c)} \bar{\phi}^{(c) \ast}(\mathbf{r}) \frac{\partial^{2}}{\partial x^{2}} \bar{\phi}^{(c)}(\mathbf{r}) + \sum_{i} N^{(s)}_{i} \bar{\phi}^{(s) \ast}_{i}(\mathbf{r}) \frac{\partial^{2}}{\partial x^{2}} \bar{\phi}^{(s)}_{i}(\mathbf{r}) \right).
\end{align}
By imposing the energy variation $\delta E_0 = \delta E_{k} + \delta E_{2B} + \delta E_{3B}$ to vanish at first order in $\lambda$ we can find
\begin{align}
	2 E^{x}_{k} + E_{2B} + 2 E_{3B} = 0.
\end{align}
Analogous expressions can be obtained by impose same scaling transform in $y$ directions. By summing over the $x$ and the $y$ directions, we can find the virial relation
\begin{align}\label{virial}
	E_{k} + E_{2B} + 2 E_{3B} = 0
\end{align}
for our two-dimensional system.

In Fig.~\ref{energy}, we plot the $N$ dependence of the energies, where $\varepsilon_{k}=E_{k}/N$, $\varepsilon_{2B}=E_{2B}/N$, and $\varepsilon_{3B}=E_{3B}/N$. As can be seen, unlike $\varepsilon_{2B}$ and $\varepsilon_{3B}$ which are monotonic functions of $N$, the variation of $\varepsilon_{k}$ is negatively correlated to that of the radial size $\sigma_{\rm GST}$ because $\varepsilon_{k}\propto 1/\sigma_{\rm GST}^2$. Furthermore, the phase transitions can also be visualized from the behavior of the energies. In particular, it appears that the SVS-to-SCS transition occurs when $\varepsilon_k=\varepsilon_{3B}$, which will be proven analytically in Sec.~\ref{sec3rd}. 

\section{density profile}\label{secdens}

\begin{figure}[ptb]
\centering
\includegraphics[width=0.8\columnwidth]{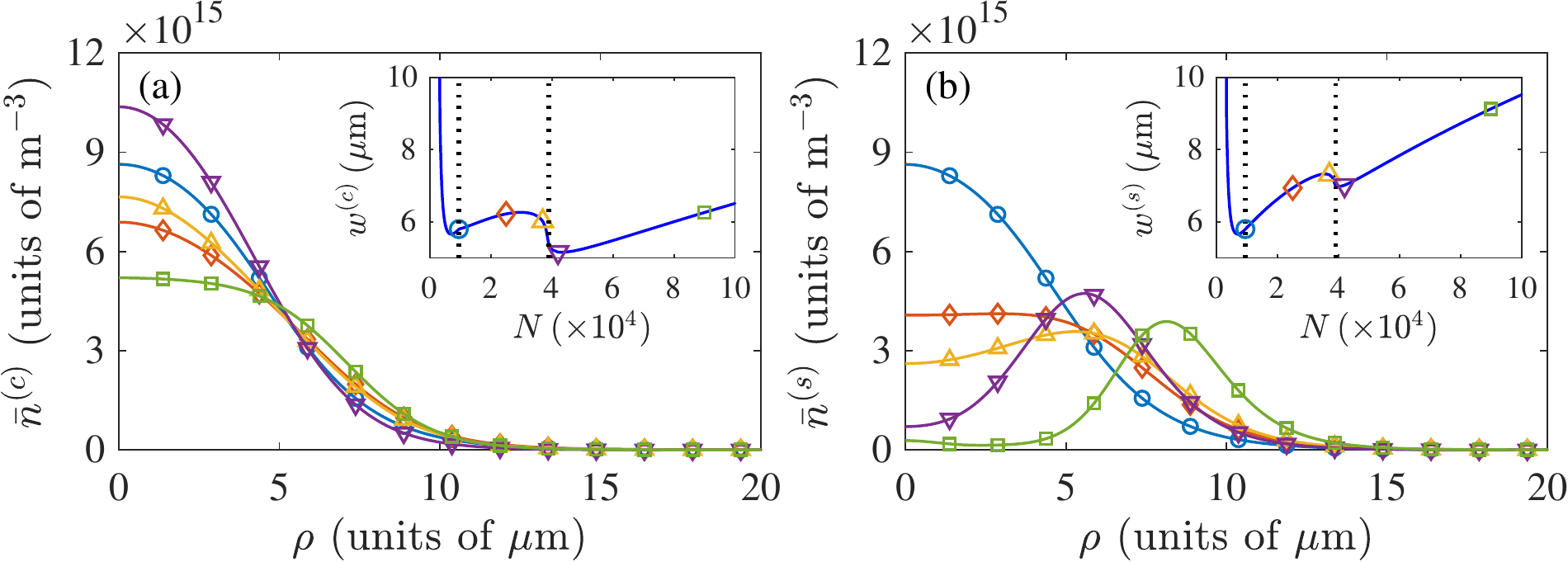}
\caption{(color online). Normalized density profiles of the coherent (a) and the squeezed atoms (b) with $\delta a=-3.062 a_{0}$ and $N=9.5\times 10^3$ ($\ocircle$), $2.5\times 10^4$ ($\diamondsuit$), $3.7\times 10^4$ ($\triangle$), $4.2\times 10^4$ ($\bigtriangledown$), and $9\times 10^4$ ($\square$). Insets in (a) and (b) show the $N$ dependence of $w^{(c)}$ and $w^{(s)}$, respectively. Vertical dotted lines mark the locations of two phase transitions. Symbols on $w^{(c,s)}$ mark the $N$ values used to plot the corresponding density profiles.}\label{dens}
\end{figure}

Here we present the normalized density profiles of the coherent and the squeezed atoms, i.e., $\bar n^{(c)}(\rho)=|\bar{\phi}^{(c)}(\rho)|^2$ and $\bar n^{(s)}(\rho)=n^{(s)}(\rho)/N^{(s)}$, here $\rho=\sqrt{x^2+y^2}$ and we have used the fact that both $\bar n^{(c)}$ and $\bar n^{(s)}$ are axially symmetric. For convenience, we introduce the width of the coherent/squeezed atoms $$w^{(c,s)}=\left[\int dxdy\rho^2 \bar n^{(c.s)}(x,y)\right]^{1/2}$$ to characterize the geometry of the states. Clearly, $w^{(c,s)}$ are better defined than the radial size when $\bar n^{(c,s)}(\rho)$ deviate significantly from Gaussian function.

In Fig.~\ref{dens}, we plot $\bar n^{(c,s)}(\rho)$ for various $N$'s in SCS and CS phases. Immediately to the right of the SVS-to-SCS transition, e.g. $N=9.5\times 10^3$, it is found that $\bar n^{(c)}$ is identical to $\bar n^{(s)}$. This observation is understandable as the coherent atoms originates from the three-body repulsion which is simply proportional to $\bar n^{(s)}$ when $N^{(c)}$ is essentially zero. When $N$ is further increased, $\bar n^{(c)}(\rho)$ changes in according with the size of the coherent part as shown in Fig.~\ref{dens}(a). Particularly, in the large $N$ limit, the top of $\bar n^{(c)}(\rho)$ becomes flatted and significantly deviates from a Gaussian function, which is in consistent to the liquid nature of the condensate. As to the profiles of the squeezed atoms, $\bar n^{(s)}(\rho=0)$ continuously decreases with $N$ such that $\bar n^{(s)}(\rho)$ is no longer a monotonic decreasing function of $\rho$ in the CS phase, which further demonstrates the squeezing here is quantum depletion due to the three-body repulsion.

\section{SVS-to-SCS transition}\label{sec3rd}
Here we derive the equation describing the boundary between the SVS and the SCS phases. To this end, we note that the 2B and 3B interaction energies for a SCS are, respectively,
\begin{align}
	E_{2B} &= \frac{4\pi\hbar^2 \delta a}{M}\frac{\sqrt{a_{\uparrow\uparrow} a_{\downarrow\downarrow}}}{(\sqrt{a_{\uparrow\uparrow}} + \sqrt{a_{\downarrow\downarrow}})^{2}}\int d\mathbf{r} \left[ N^{(c) 2} | \bar{\phi}^{(c)}(\mathbf{r}) |^{4} + 6 N^{(c)} N^{(s)} | \bar{\phi}^{(c)}(\mathbf{r}) |^{2} | \bar{\phi}^{(s)}(\mathbf{r}) |^{2} + 3 N^{(s) 2} | \bar{\phi}^{(s)}(\mathbf{r}) |^{4} \right], \notag \\
	E_{3B} &= \frac{g_{3}}{3!} \int d\mathbf{r} \left[ N^{(c) 3} | \bar{\phi}^{(c)}(\mathbf{r}) |^{6} + 15 N^{(c) 2} N^{(s)} | \bar{\phi}^{(c)}(\mathbf{r}) |^{4} | \bar{\phi}^{(s)}(\mathbf{r}) |^{2} + 45 N^{(c)} N^{(s) 2} | \bar{\phi}^{(c)}(\mathbf{r}) |^{2} | \bar{\phi}^{(s)}(\mathbf{r}) |^{4} \right.\notag \\
	&\quad\qquad\qquad\left.+ 15 N^{(s) 3} | \bar{\phi}^{(s)}(\mathbf{r}) |^{6} \right].
\end{align}
In the vicinity of the SVS-to-SCS transition, the squeezed and the coherent modes have the same density profile (see the main text for details), i.e., $\bar n({\mathbf r})\equiv \bar n^{(s)}({\mathbf r})=\bar n^{(c)}({\mathbf r})$. Then, the interaction energies reduce to
\begin{align}
	E_{2B} &= \omega_{2B} \left(N^{(c) 2} + 6 N^{(c)} N^{(s)} + 3N^{(s) 2}\right),\notag \\
	E_{3B} &= \omega_{3B} \left(N^{(c) 3} + 15 N^{(c) 2} N^{(s)} + 45 N^{(c)} N^{(s) 2} + 15 N^{(s) 3}\right),
\end{align}
where 
\begin{align}
	\omega_{2 \rm{B}} = \frac{4\pi\hbar^2 \delta a}{M}\frac{\sqrt{a_{\uparrow\uparrow} a_{\downarrow\downarrow}}}{(\sqrt{a_{\uparrow\uparrow}} + \sqrt{a_{\downarrow\downarrow}})^{2}} \int d\mathbf{r} \bar{n}^{2}(\mathbf{r}) \quad\mbox{ and }\quad \omega_{3 \rm{B}} = \frac{g_{3}}{3!} \int d\mathbf{r} \bar{n}^{3}(\mathbf{r}).
\end{align}

To find the phase boundary, we consider a pure squeezed state with all $N$ atoms occupying the squeezed mode. The total energy is then
\begin{align}
E_0(N)=\varepsilon_kN+3\omega_{2B}N^2+15\omega_{3B}N^3.\label{engsqu}
\end{align}
Now, we add $\delta N$ atom to the condensate. If all newly added atoms go to the squeezed mode, we have
\begin{align}
E_0^{({\rm SVS})}(N+\delta N)=\varepsilon_k(N+\delta N)+3\omega_{2B}(N+\delta N)^2+15\omega_{3B}(N+\delta N)^3;
\end{align}
while if all newly added atoms go to the coherent mode, we have
\begin{align}
E_0^{({\rm SCS})}(N+\delta N)=\varepsilon_k(N+\delta N)+\omega_{2B}(3N^2+6N\delta N+\delta N^2)+\omega_{3B}(15N^3+45N^2\delta N+15N\delta N^2+\delta N^3).
\end{align}
If the transition happens at $N$, we must have $E_0^{({\rm SCS})}(N+\delta N)<E_0^{({\rm SVS})}(N+\delta N)$, which, to the second order in $\delta N$, leads to $$-\omega_{2B}<15\omega_{3B}N.$$ This result says that the atom number at the third order phase transition, $N_3^*$, satisfies $-\omega_{2B}=15\omega_{3B}N_3^*$. Making use of Eq.~\eqref{engsqu}, we prove $-E_{2B} = 3 E_{3B}$ and, subsequently,
$$E_{k} = E_{3B}$$
on the boundary of the SVS-to-SCS transition. Finally, combined with the virial relation and Eq.~\eqref{engsize}, we obtain
\begin{align}
N_{\rm 3rd}^*=\frac{3}{\tilde g_2^{(s)}}.
\end{align}

\section{SCS-to-CS transition}\label{sec1st}

In this section, we derived a criterion for the first-order SCS-to-CS transition based on the stability of Bogoliubov excitation in the the presence of 3B interactions. Specifically, in the CS phase, we may neglect the contributions from terms $\langle \delta \hat{\psi}^{\dagger}( \mathbf{r}^{\prime} ) \delta \hat{\psi}( \mathbf{r} ) \rangle$ and $\langle \delta\hat{\psi}( \mathbf{r}^{\prime} ) \delta \hat{\psi}( \mathbf{r} ) \rangle$ such that Eqs.~\eqref{eta}-\eqref{delta} reduce to
\begin{align}
	\eta_{\alpha} &= \left(h^{\prime}_{\alpha} + \sum_{\beta} g^{\prime}_{\alpha \beta} \vert \phi^{(c)}_{\beta}( x, y ) \vert^{2} + \frac{g^{\prime}_{3}}{2} \sum_{\beta, \gamma}\vert \phi^{(c)}_{\beta}( x, y ) \vert^{2} \vert \phi^{(c)}_{\gamma}( x, y ) \vert^{2} \right) \phi^{(c)}_{\alpha}( x, y ) \notag \\
	\mathcal{E}_{\alpha \beta} &= \left( h^{\prime}_{\alpha} + \sum_{\gamma} g^{\prime}_{\alpha \gamma} \vert \phi^{(c)}_{\gamma}( x, y ) \vert^{2} \right) \delta_{\alpha \beta} + g^{\prime}_{\alpha \beta} \phi^{(c)}_{\alpha}( x, y ) \phi^{(c) \ast}_{\beta}( x, y )\nonumber\\
	&\quad + g^{\prime}_{3} \sum_{\gamma} \left( \sum_{\gamma^{\prime}} \frac{1}{2} \vert \phi^{(c)}_{\gamma}( x, y ) \vert^{2} \vert \phi^{(c)}_{\gamma^{\prime}}( x, y ) \vert^{2} \delta_{\alpha \beta}+ \vert \phi^{(c)}_{\gamma}( x, y ) \vert^{2} \phi^{(c)}_{\alpha}( x, y ) \phi^{(c) \ast}_{\beta}( x, y )\right) \notag \\
	\Delta_{\alpha \beta} &= g^{\prime}_{\alpha \beta} \phi^{(c)}_{\alpha}( x, y ) \phi^{(c)}_{\beta}( x, y ) + g^{\prime}_{3} \sum_{\gamma} \vert \phi^{(c)}_{\gamma}( x, y ) \vert^{2} \phi^{(c)}_{\alpha}( x, y ) \phi^{(c)}_{\beta}( x, y ),
\end{align}
where $h^{\prime}_{\alpha} = - \hbar^{2} (\partial_x^2+\partial_y^2) / ( 2M ) - \mu_{\alpha}$ is the single particle term and $g_{\alpha\beta}'=g_{\alpha\beta}/(\sqrt{2\pi}a_z)$ and $g_3'=g_3/(\sqrt{3}\pi a_z^2)$ are the interaction parameters of quasi-2D system. Meanwhile, the covariance matrix $\Gamma$ reduces to a unit matrix. As a result, the steady-state condition $\partial_{\tau} \Phi = 0$ leads to the Gross-Pitaevskii equations
\begin{align}
\left( h^{\prime}_{\alpha} + \sum_{\beta} g^{\prime}_{\alpha \beta} \vert \phi^{(c)}_{\beta}( x, y ) \vert^{2} + \frac{g^{\prime}_{3}}{2} \sum_{\beta, \gamma}\vert \phi^{(c)}_{\beta}( x, y ) \vert^{2} \vert \phi^{(c)}_{\gamma}( x, y ) \vert^{2} \right) \phi^{(c)}_{\alpha}( x, y ) = 0. \label{GPEs}
\end{align}
For simplicity, we consider a homogeneous gas. The chemical potentials can then be solved to be
\begin{align}\label{chemical potential}
	\mu_{\alpha} = \sum_{\beta} g^{\prime}_{\alpha \beta} n_{\beta} + \frac{g^{\prime}_{3}}{2} \sum_{\beta, \gamma} n_{\beta} n_{\gamma},
\end{align}
where $n_{\alpha}=|\phi^{(c)}_\alpha|^2$ is the density of $\alpha$th component. 

\begin{figure}[ptb]
	\centering
	\includegraphics[width=0.5\columnwidth]{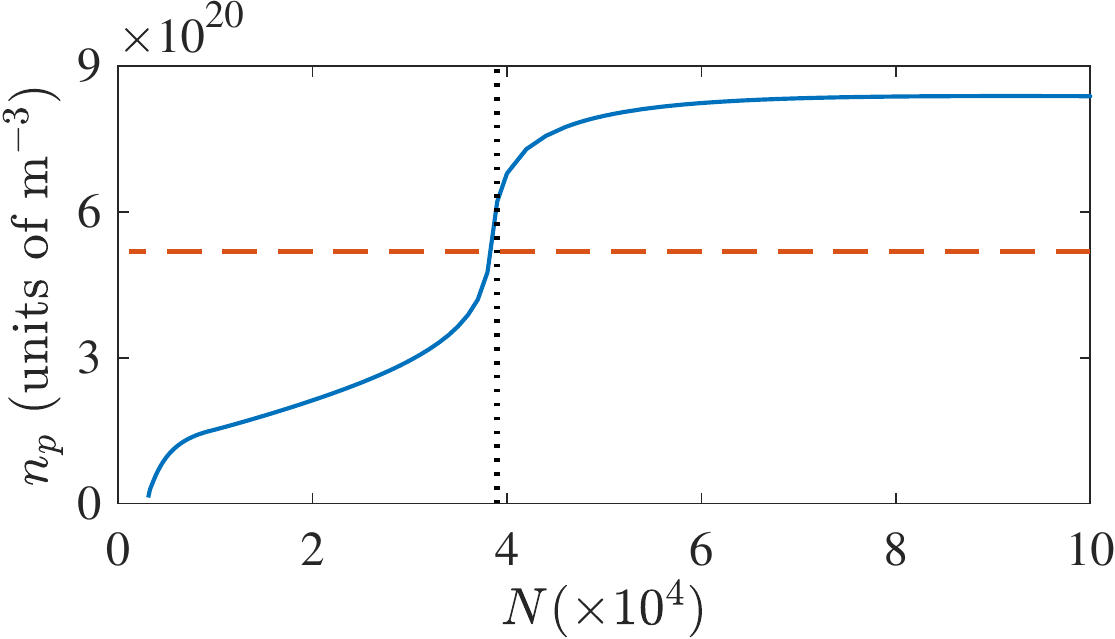}
	\caption{(color online). Peak condensate density $n_p$ as a function of atom number $N$ for $\delta a=-3.062 a_{0}$. The dashed line marks the critical condensate density $n_1^*$. The vertical dotted line denotes the boundary between the SCS and the CS phases.}
	\label{pb_c2cs}
\end{figure}

In the momentum space, the mean-field Hamiltonian becomes
\begin{align}
	H_{\mathrm{MF}}=E_{0} + \sum_{k} \frac{1}{2}: \delta \hat{\Psi}^{\dagger}_{k} \mathcal{H}_{k} \delta \hat{\Psi}_{k}:
\end{align}
where $\delta \hat{\Psi}_{k}$ is the Fourier transform of the fluctuation operator $\delta \hat{\Psi}$ and $\mathcal{H}_{k} = \begin{pmatrix}
	\mathcal{E}_{k} & \Delta_{k} \\
	\Delta^{\dagger}_{k} & \mathcal{E}^{T}_{k}
\end{pmatrix}$ with
\begin{align}
	\mathcal{E}_{k}=\begin{pmatrix}
		\epsilon_k+(g^{\prime}_{\uparrow \uparrow}+g^{\prime}_{3} \sum_{\beta} n_{\beta}) n_{\uparrow} & (g^{\prime}_{\uparrow \downarrow}+g^{\prime}_{3} \sum_{\beta} n_{\beta}) \sqrt{n_{\uparrow} n_{\downarrow}} \\
		(g^{\prime}_{\uparrow \downarrow}+g^{\prime}_{3} \sum_{\beta} n_{\beta}) \sqrt{n_{\uparrow} n_{\downarrow}} & \epsilon_k+(g^{\prime}_{\downarrow \downarrow}+g^{\prime}_{3} \sum_{\beta} n_{\beta}) n_{\downarrow}
	\end{pmatrix}
\end{align}
and
\begin{align}
	\Delta_{k}=\begin{pmatrix}
		(g^{\prime}_{\uparrow \uparrow}+g^{\prime}_{3} \sum_{\beta} n_{\beta}) n_{\uparrow} & (g^{\prime}_{\uparrow \downarrow}+g^{\prime}_{3} \sum_{\beta} n_{\beta}) \sqrt{n_{\uparrow} n_{\downarrow}} \\
		(g^{\prime}_{\uparrow \downarrow}+g^{\prime}_{3} \sum_{\beta} n_{\beta}) \sqrt{n_{\uparrow} n_{\downarrow}} & (g^{\prime}_{\downarrow, \downarrow}+g^{\prime}_{3} \sum_{\beta} n_{\beta}) n_{\downarrow}
	\end{pmatrix}.
\end{align}
Here $\epsilon_k=\hbar^2k^2/(2M)$. After diagonalizing $\mathcal{H}_k$, we obtain the Bogoliubov excitation spectra
\begin{align}
	\xi_{\pm}^{2}(k)=\frac{1}{2}\left[\xi_{\uparrow}^{2}+\xi_{\downarrow}^{2} \pm \sqrt{(\xi_{\uparrow}^{2}-\xi_{\downarrow}^{2})^{2}+16(g^{\prime}_{\uparrow \downarrow}+g^{\prime}_{3} n_{t})^{2} n_{\uparrow} n_{\downarrow} \varepsilon_{k}^{2}}\,\right]
\end{align}
where $n_t=n_\uparrow+n_\downarrow$ is the total condensate density, and
\begin{align}\label{ess}
	\xi_{\alpha}^{2}=\epsilon_{k}\left[\epsilon_{k}+2(g^{\prime}_{\alpha \alpha}+g^{\prime}_{3} n_{t}) n_{\alpha}\right]
\end{align}
are the familiar quasiparticle energies in a single-component condensate of spin-$\alpha$ atoms. Apparently, the CS phase becomes unstable when quasiparticle energy in the lower branch is imaginary, i.e.,
\begin{align}\label{stabel relation}
	(\xi_{\uparrow}^{2}+\xi_{\downarrow}^{2})^{2}<(\xi_{\uparrow}^{2}-\xi_{\downarrow}^{2})^{2}+16(g^{\prime}_{\uparrow \downarrow}+g^{\prime}_{3} n_{t})^{2} n_{\uparrow} n_{\downarrow} \epsilon^{2}_{k}.
\end{align}
After some straightforward calculations, we find
\begin{align}
n_t < n_{\rm 1st}^*\equiv \sqrt{\frac{3 \pi}{2}} \frac{a_{z}}{g_{3}} \frac{4 \pi \hbar^{2}}{M} \frac{a^{2}_{\uparrow \downarrow} - a_{\uparrow \uparrow} a_{\downarrow \downarrow}}{ ( a_{\uparrow \uparrow} + a_{\downarrow \downarrow} - 2a_{\uparrow \downarrow} )},
\end{align}
which can be interpreted as the criterion for the CS-to-SCS transition. In Fig.~\ref{pb_c2cs}, we plot the numerically obtained peak condensate density $n_p$ as a function of $N$. As can be seen, the boundary for the SCS-to-CS transition  determined with the criterion $n_p=n_{\rm 1st}^*$ is in good agreement with numerical result.

\section{Variational analysis for radial size}\label{secvariational}
Let $\bar\phi({\mathbf r})$ be a normalized mode function, the kinetic energy and the two-body interaction energies can be generally expressed as 
\begin{align}
E_{k} &= - \frac{\hbar^2N}{2 M} \int d{\mathbf r} \bar{\phi}^{\ast}({\mathbf r}) \nabla^{2} \bar{\phi}({\mathbf r}), \\
E_{2B} &= g_2' N^{2} \int d{\mathbf r} | \bar{\phi}({\mathbf r}) |^{4}. 
\end{align}
We further assume that the energy associated with the stabilization force is
\begin{align}
E_{\nu B}=g_\nu' N^{\nu}\int d{\mathbf r}|\bar{\phi}({\mathbf r})|^{2\nu}\qquad (\mbox{for $\nu>2$}).
\end{align}
To proceed, we assume that the mode function adopts a Gaussian form, i.e.,
\begin{align}
	\bar{\phi}({\mathbf r}) = \frac{1}{\sqrt{\pi}\sigma } e^{-(x^2+y^2)/ (2 \sigma ^2)}\frac{1}{(\sqrt{\pi}a_z)^{1/2}}e^{-z^2/(2a_z^2)}
\end{align}
where $a_z=\sqrt{\hbar/(M\omega_z)}$ is the width of the axial harmonic oscillator. The total energy of the self-bound droplets can then be expressed as
\begin{align}
	\varepsilon_0 = \frac{\hbar^{2}}{2M} \left(\frac{1}{\sigma^{2}} - \tilde{g}_{2} \frac{N}{\sigma^{2}} + \tilde{g}_{\nu} \frac{N^{\nu-1}}{\sigma^{2(\nu-1)}}\right).
\end{align}

\begin{figure}[ptb]
	\centering
	\includegraphics[width=0.98\columnwidth]{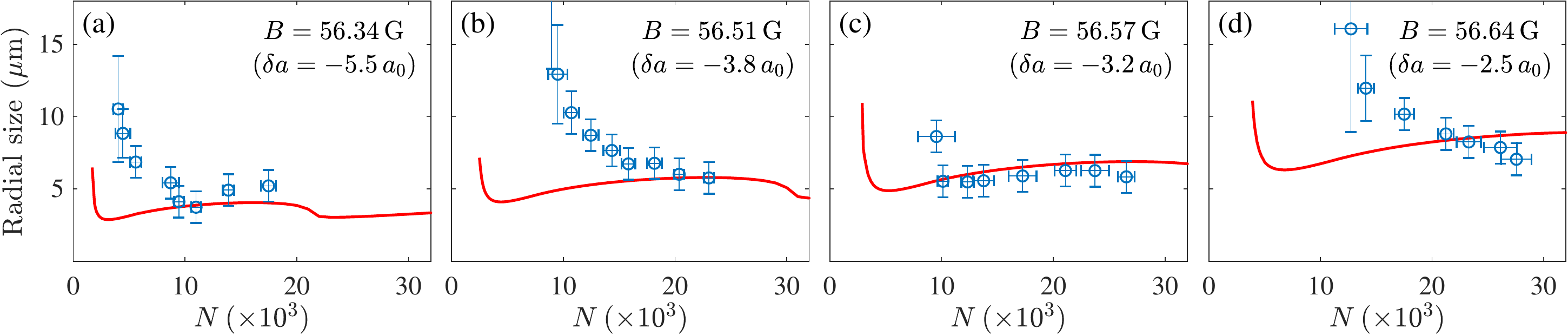}
	\caption{(color online). Radial size $\sigma $ versus atom number $N$ for various magnetic fields. Solid line is computed using the GST with $g_3=6.65 \times 10^{-39}\hbar {\rm m}^{6}/{\rm s}$. Empty circles are experimental data extracted from Ref.~\cite{Cabrera_SM}.}
	\label{pb_c2cs_2D}
\end{figure}

To derive the reduced interaction parameters $\tilde g_2^{(c,s)}$ and $\tilde g_{\nu}^{(c,s)}$, we first consider a pure coherent state for which the two-body and the thee-body interactions are
\begin{align}\label{energy apart_coh}
	E_{2B}^{(c)} &= \frac{4\pi\hbar^2 \delta a}{M}\frac{\sqrt{a_{\uparrow\uparrow} a_{\downarrow\downarrow}}}{(\sqrt{a_{\uparrow\uparrow}} + \sqrt{a_{\downarrow\downarrow}})^{2}} N^{2} \int d{\mathbf r} | \bar{\phi}({\mathbf r}) |^{4}, \notag \\
	E_{3B}^{(c)} &= \frac{g_{3}}{3!} N^{3} \int d{\mathbf r} | \bar{\phi}({\mathbf r}) |^{6}.\nonumber
\end{align}
The reduced interaction parameters can be evaluated to be
\begin{align}
\tilde{g}^{(c)}_{2} &= \frac{4}{\sqrt{2 \pi}} \frac{\sqrt{a_{\uparrow\uparrow} a_{\downarrow\downarrow}}}{(\sqrt{a_{\uparrow\uparrow}} + \sqrt{a_{\downarrow\downarrow}})^{2}} \frac{|\delta a|}{a_{z}}, \notag \\
\tilde{g}^{(c)}_{3} &= \frac{Mg_3}{9\sqrt{3} \pi^3\hbar^2a_z^2}.
\end{align}
Next, for single-mode squeezed vacuum state, we have
\begin{align}\label{energy apart}
	E_{2B}^{(s)} &= 3\frac{4\pi\hbar^2 \delta a}{M}\frac{\sqrt{a_{\uparrow\uparrow} a_{\downarrow\downarrow}}}{(\sqrt{a_{\uparrow\uparrow}} + \sqrt{a_{\downarrow\downarrow}})^{2}} N^{2} \int d{\mathbf r} | \bar{\phi}({\mathbf r}) |^{4}, \notag \\
	E_{3B}^{(s)} &= \frac{15 g_{3}}{3!} N^{3} \int d{\mathbf r} | \bar{\phi}({\mathbf r}) |^{6}.\nonumber
\end{align}
which lead to $\tilde{g}^{(s)}_{2} = 3\tilde{g}^{(c)}_{2}$ and $\tilde{g}^{(s)}_{3} = 15\tilde{g}^{(c)}_{3}$. Finally, for the EGPE theory~\cite{Petrov2015_SM}, $\tilde{g}_{5/2}^{(c)}$ can be similarly evaluated as
\begin{align}
	\tilde{g}_{5/2}^{(c)} = \frac{1024}{75 \pi} \left( \frac{\sqrt{a_{\uparrow \uparrow} a_{\downarrow \downarrow}}}{1 + \sqrt{a_{\downarrow \downarrow} / a_{\uparrow \uparrow}}} \right)^{5/2} \frac{\sqrt{2/5}}{a^{3/2}_{z} \pi^{3/4}} f\left( \frac{a^{2}_{\uparrow \downarrow}}{a_{\uparrow\uparrow} a_{\downarrow\downarrow}}, \sqrt{\frac{a_{\downarrow \downarrow}}{a_{\uparrow \uparrow}}} \right)
\end{align}
where
\begin{align}
	f(x, y) = \sum_{\eta=\pm1}\frac{1}{4\sqrt{2}}\left(1+y +\eta \sqrt{(1-y)^{2}+4 x y}\right)^{5 / 2}.
\end{align}

In Fig.~\ref{pb_c2cs_2D}, we compare, the numerically computed $\sigma_{\rm GST}$ curves with experimental data~\cite{Cabrera_SM} not shown in the main text. Apparently, some data do not have the signature V shape close to the critical atom number for the self-bound droplet. Therefore, high precision data for radial size are still needed. 

\begin{figure}[ptb]
	\centering
	\includegraphics[width=0.8\columnwidth]{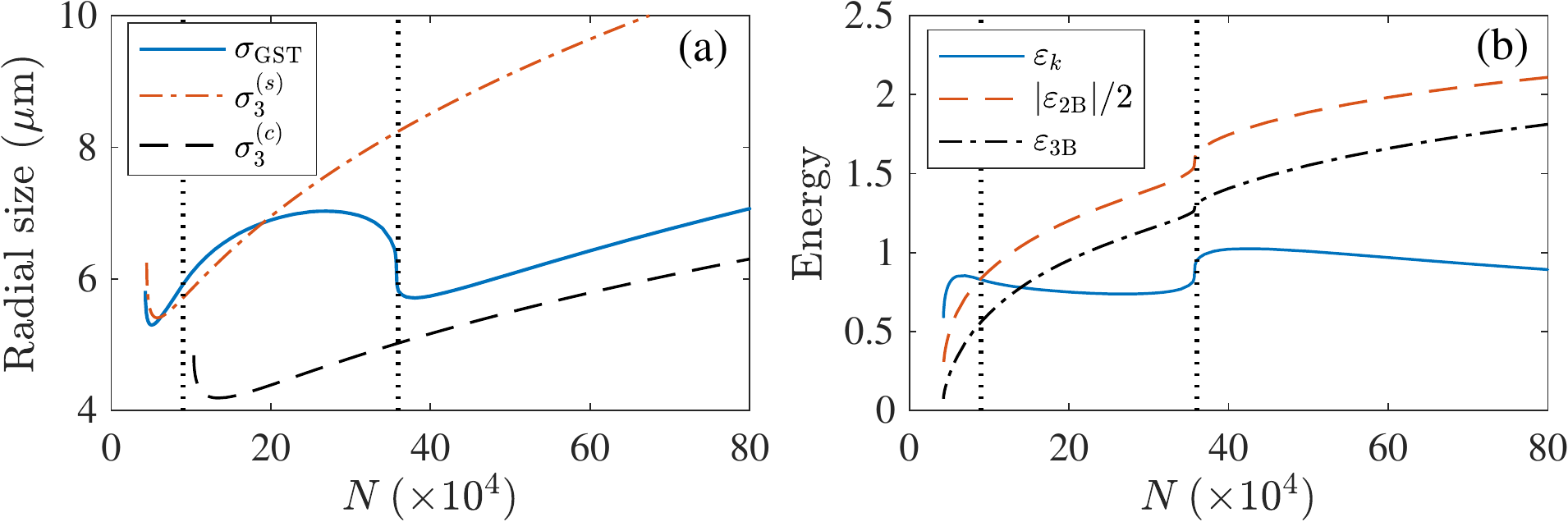}
	\caption{(color online). (a) $\sigma_r$, $\sigma_r^{(s)}$, and $\sigma_r^{(c)}$ versus $N$. (b) $\varepsilon_{k}$, $\varepsilon_{2B}$, and $\varepsilon_{3B}$ as functions of $N$. Here the $s$-wave scattering length is fixed at $\delta a=-3.062 a_{0}$ and energy is in units of $\hbar^2/(ML^2)$ with $L=6\,\mu{\rm m}$. From left to right, two vertical dotted lines mark the critical atom numbers for the third- and the first-order phase transitions, respectively.}
	\label{with3d}
\end{figure}

\section{Three-Dimensional Self-bound Droplets} \label{sec3ddrop}

For three-dimensional (3D) gases, the properties of the self-bound droplet states~\cite{Semeghini2018_SM} can be studied similarly using the Gaussian-state theory. Here we also obtain SVS, SCS, and CS phases as those found in two-dimensional droplets. Figure~\ref{with3d}(a) shows the typical behavior of the radial size $\sigma_r$ as a function atom number $N$. Apparently, $\sigma_r(N)$ also possesses a double-dip structure. In addition, the SVS-to-SCS and the SCS-to-CS transitions are of third and first order, respectively.

Analytically, we may also carry out similar analyses to those in two-dimensional gases. Here, we only present the main results for 3D binary droplets. The virial relation in 3D takes the form
\begin{align}\label{virial 3D}
	2 E_{k} + 3 E_{2B} + 6 E_{3B} = 0.
\end{align}
With the assumption of a Gaussian mode function
\begin{align}
	\bar{\phi}({\mathbf r}) = \frac{1}{\left(\sqrt{\pi}\sigma\right)^{3/2}} e^{-r^2/ (2 \sigma^2)}
\end{align}
with $\sigma_r$ being the radial size, the energy per atom can then be expressed as
\begin{align}\label{particle energy 3D}
	\varepsilon_0 =\frac{3 \hbar^{2}}{4M} \left(\frac{1}{\sigma^{2}} - \tilde{g}_{2} \frac{N}{\sigma^{3}} + \tilde{g}_{3} \frac{N^{2}}{\sigma^{6}}\right),
\end{align}
where the reduced interaction parameters depend on the quantum phases of the condensate. Specifically, for SVS, $\tilde g_2$ and $\tilde g_3$ becomes
\begin{align}
\tilde g_2^{(s)}= \frac{8}{\sqrt{2 \pi}} \frac{| \delta a |\sqrt{a_{\uparrow\uparrow} a_{\downarrow\downarrow}}}{(\sqrt{a_{\uparrow\uparrow}} + \sqrt{a_{\downarrow\downarrow}})^{2}}\quad\mbox{and}\quad 
\tilde g_3^{(s)}= \frac{10 M}{9 \sqrt{3} \pi^{3} \hbar^{2}} g_{3},
\end{align}
respectively. And similar to the 2D case, $\tilde g_2^{(c)}=\tilde g_2^{(s)}/3$ and $\tilde g_3^{(c)}=\tilde g_3^{(s)}/15$ for CS.

The equilibrium radial size is the solution of equation $\partial\varepsilon_0/\partial\sigma_r=0$ that satisfies $\partial^2\varepsilon_0/\partial\sigma_r^2>0$. In Fig.~\ref{with3d}, we plot the equilibrium radial sizes $\sigma_r^{(s)}(N)$ and  $\sigma_r^{(c)}(N)$ obtained with the reduced interactions parameters $(\tilde g_2^{(s)},\tilde g_3^{(s)})$ and $(\tilde g_2^{(c)},\tilde g_3^{(c)})$, respectively. As can be seen, they are in rough agreement with the numerical results at SVS and CS regimes. In particular, the critical atom number of the self-bound droplet is determined by the equations: $\partial\varepsilon_0/\partial\sigma_r=0$ and $\partial^2\varepsilon_0/\partial\sigma_r^2=0$, which leads to $N_{\rm cri} = 64\sqrt{\tilde{g}_{3}}/(27 \tilde{g}^{2}_{2})$. We then find two critical atom numbers
\begin{align}
	N_{\rm cri}^{(s)} = \frac{2 \pi (\sqrt{a_{\uparrow\uparrow}} + \sqrt{a_{\downarrow\downarrow}})^{4}}{81 a_{\uparrow\uparrow} a_{\downarrow\downarrow} \delta a^{2}} \sqrt{\frac{10 M g_{3}}{ \sqrt{3} \pi^{3} \hbar^{2}}}\quad\mbox{and}\quad 
N_{\rm cri}^{(c)} = \frac{2 \pi (\sqrt{a_{\uparrow\uparrow}} + \sqrt{a_{\downarrow\downarrow}})^{4}}{9 a_{\uparrow\uparrow} a_{\downarrow\downarrow} \delta a^{2}} \sqrt{\frac{2 M g_{3}}{3\sqrt{3} \pi^{3} \hbar^{2}}}
\end{align}
for SVS and CS phases, respectively.

\begin{figure}[ptb]
	\centering
	\includegraphics[width=0.5\columnwidth]{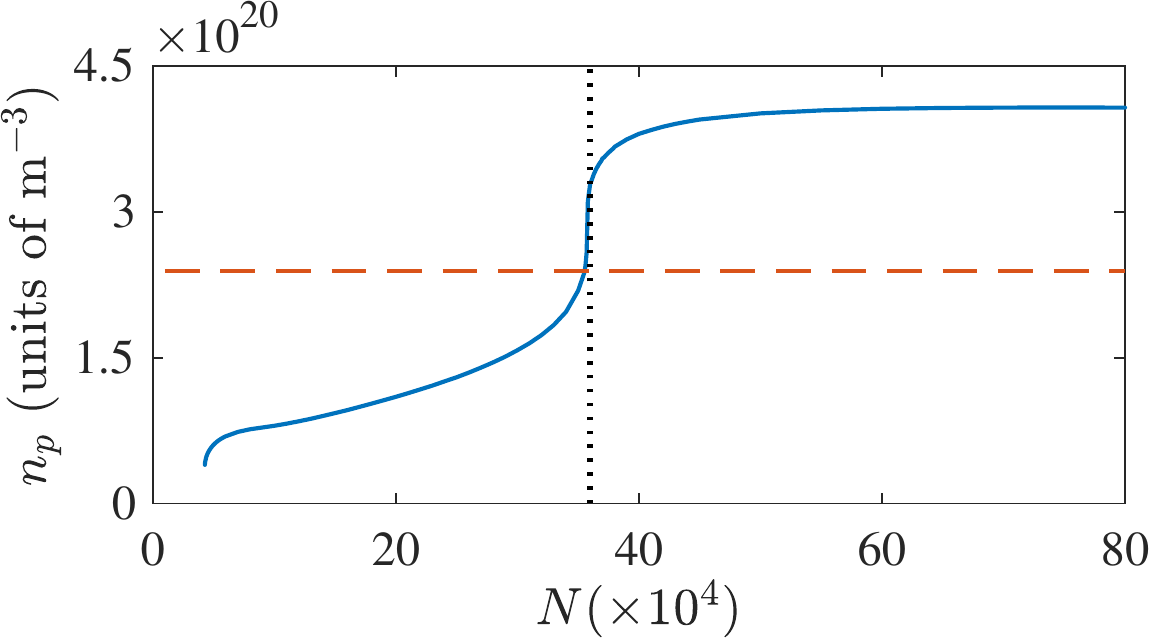}
	\caption{(color online). Peak condensate density $n_p$ as a function of atom number $N$ for $\delta a=-3.062 a_{0}$. The dashed line marks the critical condensate density $n_1^*$. The vertical dotted line denotes the boundary between the SCS and the CS phases.}
	\label{pb_c2cs_3D}
\end{figure}

In Fig.~\ref{with3d}, we plot the energy contributions as a function of $N$. It can be analytically shown that the SVS-to-SCS transition occurs when $2E_k=|E_{2B}|$ or at
\begin{align}
	N^*_{\rm 3rd} = \frac{4}{\tilde{g}^{2}_{2}} \sqrt{\frac{3 \tilde{g}_{3}}{2}} = \frac{\pi (\sqrt{a_{\uparrow\uparrow}} + \sqrt{a_{\downarrow\downarrow}})^{4}}{8 a_{\uparrow\uparrow} a_{\downarrow\downarrow} \delta a^{2}} \sqrt{\frac{5 M g_{3}}{3 \sqrt{3} \pi^{3} \hbar^{2}}}.
\end{align} 
Finally, for the SCS-to-CS transition, the phase boundary is located at roughly
\begin{align}
	n_{\rm 1st}^* = \frac{1}{g_{3}} \frac{4 \pi \hbar^{2}}{M} \frac{a^{2}_{\uparrow\downarrow} - a_{\uparrow \uparrow} a_{\downarrow \downarrow}}{( a_{\uparrow \uparrow} + a_{\downarrow\downarrow} - 2a_{\uparrow, \downarrow} )}.
\end{align}
Again, as shown in Fig.~\ref{pb_c2cs_3D}, it can be verified that the phase boundary given by the criterion $n_p=n_{\rm 1st}^*$ is in good agreement with the numerical result.

\end{document}